\documentclass[11pt]{article}
\usepackage{epsfig}
\usepackage{amsmath}

\newlength{\dinwidth}
\newlength{\dinmargin}
\title{\bf  Evolution of pomeron and odderon\\
at all angular momenta}

\author{M.A.Braun\\
Dep. of High Energy physics,
Saint-Petersburg State University,\\
198504 S.Petersburg, Russia}

\newcommand{\tdm}[1]{\mbox{\boldmath $#1$}}
\newcommand{\abar}{\bar\alpha_s}
\newcommand{\vx}{\tdm{x}}
\newcommand{\vy}{\tdm{y}}
\newcommand{\vz}{\tdm{z}}
\newcommand{\vk}{\tdm{k}}
\newcommand{\vr}{\tdm{r}}
\newcommand{\vb}{\tdm{b}}
\newcommand{\be}{\begin{equation}}
\newcommand{\ee}{\end{equation}}

\begin{document}
\maketitle

\abstract
{In the QCD the small~$x$ evolution of the interacting pomerons and odderons
 is studied with all
angular momenta $l$ taken into account.
The resulting system of coupled nonlinear evolution equations is formulated
in the momentum space and solved numerically. Excellent convergence  in $l$ is observed. Also
it is found that  states with $l>1$ play an important role and substantially reduce the
basic pomeron state at large rapidities}

\section{Introduction}

Since long ago one of the main  features of the strong interaction has been
the dominance at high energies of the C even exchange over the C odd one
("the Pomeranchuk theorem"). In the Regge language the large energy asymptotic
of the $C=+1$ amplitude is due to pomeron exchanges and that of the $C=-1$
amplitude by the odderon exchanges, both pomeron and odderon corresponding to the
leading singularities of the relevant amplitudes in the complex angular momentum plane.
 Whereas the behavior of the $C=+1$
transitions is more or less confirmed by experiments, which show the growing cross-sections,
the $C=-1$ behavior has been somewhat elusive up to this date. Since the pioneering work
~\cite{nicolescu} this behavior has been attributed to the odderon (whether a pole or not
in the complex angular momentum plane) with the intercept close or exactly equal to one.
However the experimental evidence of its existence remains inconclusive in spite of many
assertions ~\cite{ppodd,ref1,ref2,ref3,ref4}. Remarkably on  the theoretical level the existence of the odderon
has been well predicted within the QCD paradigm. In this picture the odderon appears as an object made of three
reggeized gluons ("reggeons") in the $d$-color state, as opposed to the pomeron made of two reggeons in the
colorless state. Moreover within  the Regge kinematics ( fixed $t$, $s/t\to\infty$) and in the leading approximation
in $\alpha_s\ln s$  both the pomeron and odderon leading intercepts has been found to be $1+\Delta$ ~\cite{bfkl} and exactly unity
~\cite{blv}. Here $\Delta>1$ is the well-known BFKL intercept, which predicted the growth of  cross-sections
in the strong interaction at high energies.

This picture hints to some reasons for the  weakness of the odderon exchange.
The flatness of the corresponding cross-sections as compared to the rising ones for the
pomeron exchange already make its observation very difficult. Also an extra power of $\alpha_s$
related to its three reggeon components instead of two in the pomeron presumably make its coupling to the hadrons weaker.

Many theoretical estimates   of the cross sections
for various odderon mediated
processes~\cite{etac,ryskin,ggeta,heid,brodsky,bbcv,twopi} confirm this weakness and predict small
cross sections, below the sensitivity of current experiments.  The only
exception, for which some evidence of the odderon contribution was probably
measured, is the elastic $pp$ and $p\bar{p}$ scattering at non-zero momentum transfer
~\cite{ppodd,ref1,ref2,ref3,ref4}. However the final conclusions
 from these experimental observations remain not too convincing up to now.

As is well known, in the QCD
the perturbative small~$x$ evolution equation of the pomeron amplitudes,
taking into account non-linear unitary corrections  was derived by
Balitski ~\cite{bal} and Kovchegov ~\cite{kov}. In the diagrammatic
language, the Balitski-Kovchegov (BK) equation resums BFKL pomeron fan
diagrams in the large $N_c$ limit. The BK equation may be also obtained
as the mean-field limit of the effective theory of small~$x$ gluons in the
hadron wave function (the Color Glass Condensate approach~\cite{jimwalk}).

As mentioned, in the QCD the odderon
consists of three $t$-channel reggeons  in the color singlet  $d$-state.
Generally these three gluons may occupy three different spatial points.
The small~$x$ evolution equation of this  odderon (the BKP equation)
was derived long time ago~\cite{bartels,kp}. Its leading intercept was found to lie below unity
~\cite{jw}  meaning that $C$-odd cross-sections should
decrease with energy. However later a new odderon solution was discovered
~\cite{blv} with two of the three reggeons located at the same spatial point
This "degenerated" BLV odderon has its intercept equal to exactly unity, so that its contribution to the
high-energy $C$-odd cross-sections is dominating. With the two reggeons fused into one the wave function
effectively coincides with the pomeron wave function with the negative spatial symmetry. So its small-$x$
evolution is described by the equation analogous to the BK equation for the pomerons ~\cite{ksw}.
Under some approximations this equation has recently been solved ~\cite{conlevin} where fast decrease of the odderon amplitude
at large rapidity has been  found.

Long ago the theory predicted  that the pomeron may split into two odderons ~\cite{barew}.
Therefore in the course of  small-$x$ evolution the pomeron fan diagrams may generate pairs of odderons, so that the
pomeron and odderon evolutions are interrelated.  The system of coupled  non-linear equations
involving both the $C$-even and $C$-odd amplitudes was derived in~\cite{hiim}.
This system was studied in ~\cite{motyka} under some important approximations: the translational invariance and
the lowest angular momenta  $l=0$ for the pomeron and $l=1$ for the odderon. The first approximation compelled to
substitute the odderon contribution to the pomeron to its average over the angle.

In this study we retain the first approximation (translational invariance) but give up the second to
study  evolution at all angular momenta.


\section{Formalism.}

Let $ N(\vx,\vy;\tau)$ be the pomeron density in  the transverse position plane for the collision off a large nucleus and
$O (\vx,\vz;\tau)$ be the similar density of the BLV odderon.
Let also the rapidity be $\tau=\log(1/x)$. The $\tau$-evolution of the $C$-even amplitude $ N(\vx,\vy;\tau)$ and $C$-odd amplitude
$O (\vx,\vz;\tau)$ in the leading order approximation  in powers of $\alpha_s\tau$
is described by
a system of equations~\cite{hiim}:
\[
{\partial N(\vx,\vy;\tau) \over \partial \tau} =
{\abar \over 2\pi} \int d^2 z \;
{ (\vx - \vy)^2 \over (\vx - \vz)^2  (\vz - \vy)^2}
\left[
N(\vx,\vz;\tau) + N(\vz,\vy;\tau) - N(\vx,\vy;\tau) \right.
\]
\be
\left.
- N(\vx,\vz;\tau)N(\vz,\vy;\tau)  +
O (\vx,\vz;\tau) O(\vz,\vy;\tau)
\right],
\label{pom}
\ee
\[
{\partial O(\vx,\vy;\tau) \over \partial \tau} =
{\abar \over 2\pi} \int d^2 z \;
{ (\vx - \vy)^2 \over (\vx - \vz)^2  (\vz - \vy)^2}
\left[
O(\vx,\vz;\tau) + O(\vz,\vy;\tau) - O(\vx,\vy;\tau)
\right.
\]
\be
\left.
- O(\vx,\vz;\tau)N(\vz,\vy;\tau)  -
N (\vx,\vz;\tau) O(\vz,\vy;\tau)
\right],
\label{odd}
\ee
In fact $\vx$, $\vy$ and $\vz$ represent positions of
the end points in the transverse plane of color dipoles, which interact with a large target.
The pomeron and the odderon  amplitudes have definite
parities with respect to exchange of the gluon positions, that is:
\be
N(\vy,\vx;\tau) =  N(\vx,\vy;\tau), \qquad O(\vy,\vx;\tau) =  -O(\vx,\vy;\tau).
\label{parity}
\ee
If we separate the central-of-mass (c.m) coordinate $\vb=(\vx+\vy)/2$ the amplitudes become
$N(\vb,\vy-\vx,\tau)$ and $O(\vb,\vy-\vx,\tau)$. In the large nucleus at rest
the individual nucleons interact with a very small transverse momentum transfer,
of the order of $1/R_A$ where $R_A$ is the nuclear radius. So all transverse momentum transfers
along the pomeron fan diagram result to be of the same small order and can be taken as zero.
In this case  the $C$-even amplitude  $N$ can be taken in the forward direction, which means that the
impact parameter $\vb$ is not changed in the evolution and enters  only as an external parameter.
In particular at the start of the evolution with a symmetric nuclear target one can take
$N(\vb,\vy-\vx,\tau=0)=N_b((\vy-\vx)^2,\tau=0)$. Of course this form satisfies condition
(\ref{parity}).

Inclusion of the odderon radically changes the situation. The requirement of antisymmetry in $\vx$ and $\vy$
implies that the amplitude has to depend on $(\vb,\vx-\vy)$ and be antisymmetric in this argument. Integration
over $\vb$ then gives zero, which means that the amplitude vanishes at zero momentum transfer. In principle this implies that
one has to consider the amplitude at finite momentum transfers. Apart from difficulties
for application to  collisions with a large nucleus this leads to the necessity to study evolution equations in the whole
space of two independent variables $\vx$ and $\vy$, both changing in the course of evolution.
Having mostly  in mind to study the influence of the odderon on the evolution we shall try to simplify the problem
following the idea of ~\cite{motyka}. Consider the first term on the right-hand side of Eq. (\ref{pom}) in variables with the extracted
c.m. coordinate. It has the form
\[
 {\abar \over 2\pi} \int d^2 z \;
{ (\vx - \vy)^2 \over (\vx - \vz)^2  (\vz - \vy)^2}
N(\vb',\vx-\vz;\tau)\]
where $\vb'=(\vx+\vz)/2$ is the evolved c.m. coordinate. Using $\vx=\vb+(\vx-\vy)/2$
we have $\vb'-\vb=(\vx+\vz)/2-(\vx+\vy)/2=(\vz-\vy/2$. So in the course of evolution the
impact parameter changes by the order of the average dipole dimension. If we take the
impact parameter very large as compared to the average dipole dimension then one can neglect this
change so that $\vb$ also becomes a fixed external parameter for the evolution. However in this case
it is a fixed  vector parameter with not only the magnitude of $b$ but also its direction as evolution parameter.
Then both $N$ and $O$ will depend on the vector $\vx-\vy$ in the presence of the external direction given by the
fixed $\vb$:
Thus, we assume
\be
N(\vx,\vy;\tau) = N_{\vb}(\vy-\vx,\tau), \qquad O(\vx,\vy;\tau) = O_{\vb}(\vy-\vx,\tau).
\label{approx}
\ee
having in mind to find the influence of the odderon in Eqs. (\ref{pom}) and (\ref{odd})
and leaving aside  its relevance for the actual physical case of the collision with a large nucleus.

To simplify the non-linear terms we pass to the momentum space.
and define the momentum dependent functions $\Phi(\vk,\tau)$ and
$\Psi(\vk,\tau)$
describing the pomeron and the odderon dipole densities respectively
\be
\Phi(\vk,\tau) = \int {d^2 \vr \over 2 \pi r^2} N(\vr,\tau) \exp(-i \vk\vr),\ \
\Psi(\vk,\tau) = \int {d^2 \vr \over 2 \pi r^2} O(\vr,\tau) \exp(-i \vk\vr).
\ee
From Eqs. (\ref{pom}) and (\ref{odd}) with the amplitudes having the  forms (\ref{approx})
one obtains a system of equations ~\cite{motyka}
\be
{\partial \Phi(\vk,\tau) \over \partial \tau} =
\abar\, (K\otimes \Phi)(\vk,\tau) -  \abar \Phi^2(\vk,\tau) + \abar \Psi^2(\vk,\tau),
\label{kpom}
\ee
\be
{\partial \Psi(\vk,\tau) \over \partial \tau} = \abar\, (K\otimes \Psi)(\vk,\tau) -
2 \abar \Phi(\vk,\tau)\Psi(\vk,\tau),
\label{kodd}
\ee
where the linear terms describe the standard BFKL evolution in the forward direction
\be
(K\otimes \Phi)(\vk,\tau) = \frac{1}{\pi}\int { d^2 \vk' \over (\vk - \vk')^2} \,
\left[
\Phi(\vk',\tau) - {k^2\Phi(\vk,\tau) \over
\vk'^2 + (\vk - \vk')^2}
\right]
\ee
and similar for $\Psi$

In the presence of the external direction we
develop both $\Phi$ and $\Psi$ in angular momenta $l$
\be
\Phi(\vk,\tau) = \sum_{l, even}\Phi_l(k,\tau)e^{il\varphi},\ \  \qquad
\Psi(\vk,\tau) =
\sum_{l,odd}\Psi_l(k,\tau)e^{il\varphi},
\ee
Here $\varphi$ is the angle between $\vk$ and the fixed  direction in the
transverse plane. The angular momentum $l$ goes from $\infty$ to $+\infty$
but the parity condition (\ref{parity}) requires even  angular momenta for $\Phi$ and odd ones for $\Psi$.

We obtain the following system of coupled equations for $\Phi_l$ and $\Psi_l$
\[
{\partial \Phi_l(k,\tau) \over \partial w} =
\int_0 ^\infty dk'^2\Big\{\Phi_l(k',\tau) \Big(\frac{k_<^2}{k_>^2}\Big)^{|l|}\,\frac{1}{k_>^2-k_<^2|}
 - \Phi_l(k,\tau)\frac{k^2}{{k'}^2}\Big(\frac{1}{k_>^2-k_<^2|}-\frac{1}{\sqrt{4{k'}^4+k^4}}\Big)\Big\}\]\be
- \sum_{m}\Phi_m(k,\tau)\Phi_{l-m}(k,\tau) +\abar\sum_{m}\Psi_m(k,\tau)\Psi_{l-m}(k,\tau)
\label{eqphi}
\ee
with $l$ even and
\[
{\partial \Psi_l(k,\tau) \over \partial w} =
\int_0 ^\infty dk'^2
\int_0 ^\infty dk'^2\Big\{\Psi_l(k',\tau) \Big(\frac{k_<^2}{k_>^2}\Big)^{|l|}\,\frac{1}{k_>^2-k_<^2|}
 - \Psi_l(k,\tau)\frac{k^2}{{k'}^2}\Big(\frac{1}{k_>^2-k_<^2|}-\frac{1}{\sqrt{4{k'}^4+k^4}}\Big)\Big\}\]\be
-2\sum_{m}\Phi_m(k,\tau)\Psi_{l-m}(k,\tau)
\label{eqpsi}
\ee
with $l$ odd.
Here $k_< = \min (k,k')$ and $k_> = \max (k,k')$. We also introduce the rescaled rapidity $w=\abar\tau$.
Of course $\Phi_n=0$ for $l$ odd and $\Psi_l=0$ for $n$ even. Also $\Phi_l=\Phi_{-l}$ and $\Psi_l=\Psi_{-l}$.
 So one can rewrite
\be
 \sum_{m}\Phi_m\Phi_{l-m}=\Phi_0\Phi_{|l|}+\sum_{m=1}^\infty \Phi_m (\Phi_{|l-m|}+\Phi_{|l+m|}),\ee
 \be
 \sum_{m}\Psi_m\Psi_{l-m}=\Psi_0\Psi_{|l|}+\sum_{m=1}^\infty \Psi_m (\Psi_{|l-m|}+\Psi_{|l+m|})\ee
 and
 \be
 \sum_{m}\Phi_m\Psi_{l-m}=\Phi_0\Psi_{|l|}+\sum_{m=1}^\infty \Phi_m (\Psi_{|l-m|}+\Psi_{|l+m|})
 =\Psi_0\Phi_{|l|}+\sum_{m=1}^\infty \Psi_m (\Phi_{|l-m|}+\Phi_{|l+m|}).
 \ee
These equations form the basis of our numerical
calculations

Since $\Phi_l=0$ for $l$ odd and $\Psi_l=0$ for $l$ even it is convenient to introduce
\be
\phi_l\equiv\Phi_{2l},\ \ \psi_l\equiv \Psi_{2l+1}\ee
Then both $\phi_l$ and $\psi_l$ are different from zero for all $l=0,1,...$.
In terms of $\phi_l$ and $\psi_l$
\be
\Phi=\phi_0+2\sum_{l=1}\phi_l\cos 2l\phi,\ \ \Psi=2\sum_{l=0}\psi_l\cos(2l+1)\phi.
\ee
In the following we denote the angular momentum of the pomeron as $L=2l$ and of the odderon
as $L=2l+1$.

The sums in our equations are then transformed as follows
\be
C^{PP}_l=\sum_{m=-\infty, even}^{+\infty} \Phi_m\Phi_{2l-m}=\phi_0\phi_l+\sum_{m=1}\phi_m(\phi_{|l-m|}+\phi_{l+m}),
\label{cpp}
\ee
\be
C^{OO}_l\sum_{m=-\infty, odd}^{+\infty} \Psi_m\Psi_{2l-m}=\sum_{m=0}\psi_m(\psi_{m'}+\psi_{l+m})
\label{coo}
\ee
where $m'=(|2l-2m-1|-1)/2$.
Finally
\be
C^{PO}_l=\sum_{m=-\infty, odd}^{+\infty} \Psi_m\Phi_{2l+1-m}=\sum_{m=0}\psi_m(\phi_{|l-m|}+\phi_{l+m+1}).
\label{cpo}
\ee

We recall that $\phi_0$ is directly related to the non-integrated gluon density in the nucleus.
In our normalization (see ~\cite{bra})
\be
\frac{\partial xG(x,k^2)}{\partial k^2}=\frac{N_c^2}{2\pi^3\bar{\alpha}}k^2\nabla_k^2\phi_0\Big(\ln\frac{1}{x},k^2\Big).
\label{gluden}
\ee

\section{Calculations}
\subsection{Passing to a grid in $\ln k^2$}

We pass to a logarithmic variable
$t=\ln k^2 =\ln q$ with $q=k^2$ (the units in which $q$ is measured is inferred from the initial functions for evolution).
In variable $t$ the non-integrated gluon density is
\be
\frac{\partial xG(x,k^2)}{\partial k^2}=\frac{N_c^2}{2\pi^3\bar{\alpha}}\partial_t^2\phi_0(w,t)
\label{gluden1}
\ee
where $w=\abar\ln\frac{1}{x}$.

We introduce a grid in $t$.
\be
t_i=t_{min}+id,\ \ i=0,1,...n,\ \ d=\frac{t_{max}-t_{min}}{n}.
\label{ti}
\ee
At the grid points $q_i=\exp(t_i)$ and the  fields are  $\phi_{li}=\phi_l(q_i)$ and $\psi_{li}=\psi_l(q_i)$.
We approximate the integrals over $t$ by finite sums
\be
\int_{-\infty}^{\infty}dt\,F(t)\simeq\sum_{i=1}^{n}w_{i}F(t_{i})
\label{apprint}
\ee
with points $t_{i}$ and weights $w_{i}$ depending on the chosen
approximation scheme.

Then
Eqs. (\ref{eqphi}) and (\ref{eqpsi}) pass into the system of finite matrix equations
for evolution in the rescaled rapidity $w$
\be
\frac{ d \phi_{li}}{dw}=
\sum_{j\neq i}q_j\Big(B^P_{lij}\phi_{lj}-A_{ij}\phi_{li}\Big)+\frac{\phi_{li}}{\sqrt{5}}-C^{PP}_{li}+C^{OO}_{li}
\label{eqp}
\ee
and
\be
\frac{ d \psi_{li}}{dw}=
\sum_{j\neq i}q_j\Big(B^O_{lij}\psi_{lj}-A_{ij}\psi_{li}\Big)+\frac{\psi_{li}}{\sqrt{5}}-2C^{PO}_{li}.
\label{eqo}
\ee
In these equations
\be
B^P_{lij}=\Big(\frac{q_<}{q>}\Big)^{2l}\frac{1}{|q_i-q_j|},\ \
B^O_{lij}=\Big(\frac{q_<}{q>}\Big)^{2l+1}\frac{1}{|q_i-q_j|},\ \
A_{ij}=\frac{qi}{qj}\frac{1}{\sqrt{qi^2+4q_j^2}}
\label{ab}
\ee
and terms $C^{PP}$, $C^{OO}$ and $C^{PO}$ are the nonlinear terms (\ref{cpp}), (\ref{coo}) and (\ref{cpo}).

\subsection{Initial conditions}
We assume that the target initially interacts  with  the pomeron and odderon only at the  lowest
orbital momenta $L=0$ for the pomeron and $L=1$ for the odderon. Higher orbital momenta appear only as a result of evolution.
The pomeron function is related to the forward scattering of a dipole on a  large nucleus and its initial function
can be taken in the standard manner, as in a numerous previous calculations.
A popular choice of this initial function follows the form proposed in ~\cite{GBW}. In the coordinate space
\be
N(\tau=0,{\bf x})=1-\exp\Big(-\frac{1}{4}Q_A^2x^2\Big)
\label{gbw}
\ee
where $Q_A^2=cA^{1/3}Q_1^2$ and $Q_1$ and $c$ are constants  determined by the data. From  ~\cite{dusling}
we have $Q_1^2=0.24$ GeV$^2$ and $c\simeq 0.25$.

Fourier transformation to the momentum space gives
\be
\phi_0(\tau=0,q)=-\frac{1}{2}{\rm Ei}\Big(-\frac{q}{Q_A}\Big).
\label{phi0}
\ee
To see this one can use the  Fourier transform
\be
\int\frac{d^2k}{2\pi}{\rm Ei}(-k^2)e^{i{\bf kr}}=\int_0^\infty kdk {\rm Ei}(-k^2){\rm J}_0(kr)
\ee
and the standard formula (2.12.47.8) from ~\cite{prud}
at $n=0$
\be
\int_0^\infty kdk{\rm Ei}(-bk^2){\rm J}_0(ck)=\frac{2}{c^2}\Big[1-\exp\Big(-\frac{c^2}{4b}\Big)\Big].
\ee
The inverse Fourier transform gives (\ref{phi0}).

As to the odderon initial function, in absence of the underlying clear physical picture and motivated mostly
by our desire to study the influence of the inclusion of the odderon in the evolution, we take its initial function $\psi_1(\tau=0,q)$
in the same form (\ref{gbw}) with a scaling factor $g_O$ which may take into account a possible weakness of the
odderon coupling. In fact in our numerical calculations we take $g_O=1$ to investigate qualitatively the odderon influence with
a coupling of a similar strength. Diminishing of $g_O$ will inevitably make this influence weaker.

\section{Numerical results}
We present our numerical results for $\phi$ and $\psi$ at different scaled rapidities
$w=\bar{\alpha}_s\tau$ rising from 0 to 7. With $\alpha_s\sim 0.2$ this corresponds
to natural tapidities up to $\sim 35$. As the momentum variable we choose a dimensionless
variable
\[x=\frac{t-t_{min}}{t_{max}-t_{min}}=\frac{\ln(q/q_{max})}{\ln(q_{max}/q_{min}},\ \ 0<x<1\]
with $q_{min,max}=Q_A\exp(t_{min,max}/2)$. In our calculations we chose
$t_{min}=-20$ and $t_{max}=60$, and $n=800$, which proved to be  values sufficient for a reasonable
precision ($\leq 0.0001$) at $t>-12$, that is at $Q^2> 1.5e-6$ GeV$^2$. With these  $t_{min}$ and $t_{max}$ values $x=0$ and $x=1$ correspond
to 4.95e-10 GeV$^2$ and $2.74e+25$ GeV$^2$ respectively. The typical momentum squared
10 (Gev)$^2$ corresponds to $x\simeq 0.3$.
For some cases apart from $\phi$ and $\psi$ we present the rescaled non-integrated gluon density
\[
g(w,k^2)=k^2\nabla_k^2\phi(w,k)=\partial_t^2\phi(w,t)\]
and a similar function for $\psi$ (although the physical interpretation of the latter is
somewhat obscure)

\subsection {The Pomeron}
\begin{figure}
\begin{center}
\epsfig{file=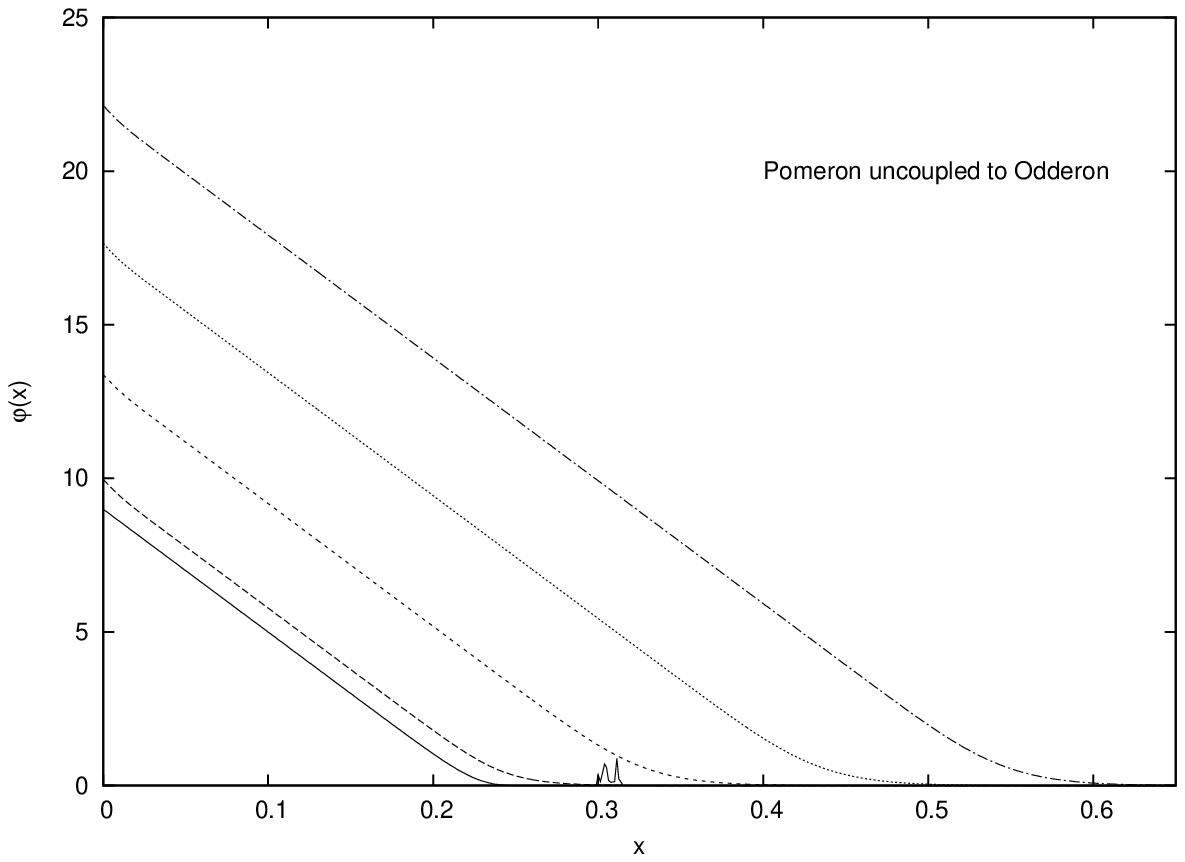, width=0.45\columnwidth }
\epsfig{file=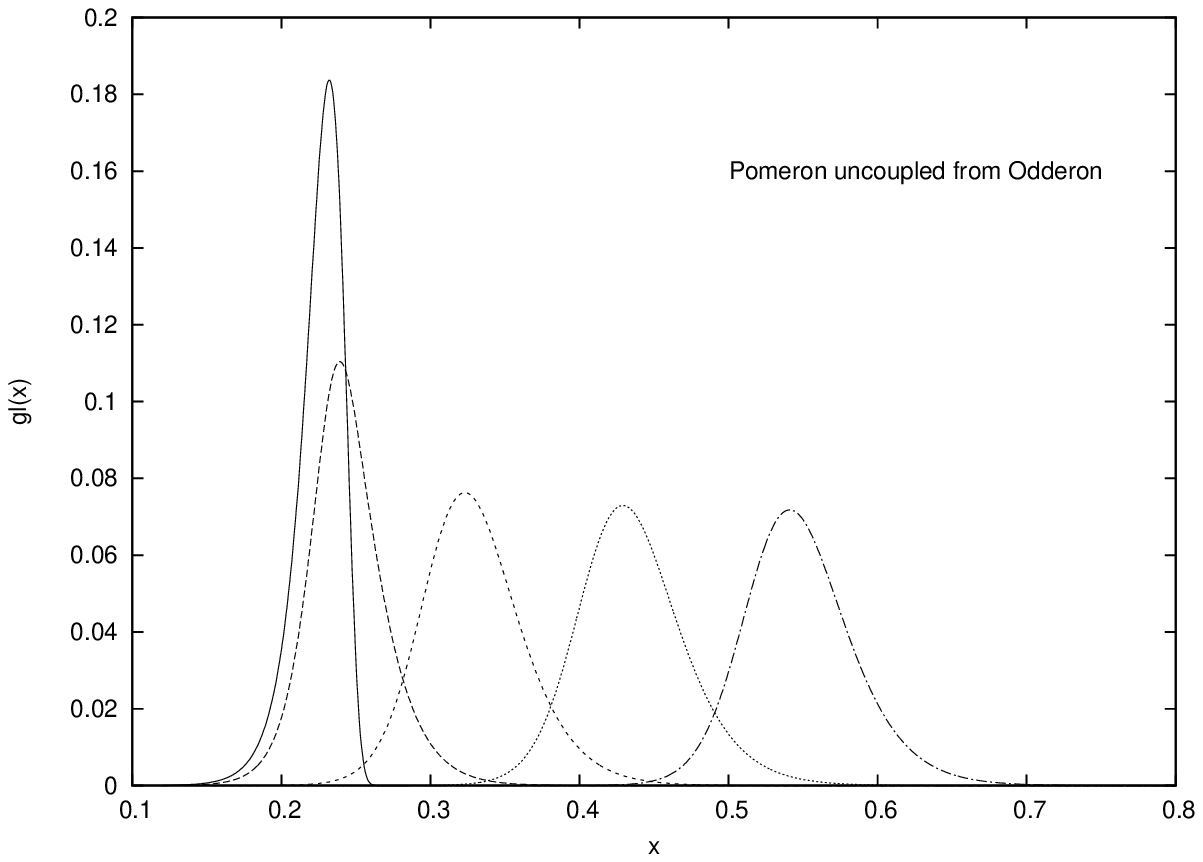, width=0.45\columnwidth }
\end{center}
\caption{The Pomeron $\phi_0$ decoupled from the odderon (left panel) and the corresponding gluon density
(right panel). Curves from top to bottom in the left panel and from left to right in the right panel
correspond to $w=0,1,3,5$ and 7 }
\label{fig1}
\end{figure}

For comparison we start with the well studied case of the pure pomeron evolution without coupling to the odderon.
This corresponds to non-linear terms $C^{OO}=C^{PO}=0$
Since we assume that only $\phi_0$ is initially coupled with the target, all $\phi_l$ with $l>0$ remain zero after
evolution and $\phi_0$ evolves according to the standard BK equation. The values of $\phi_0(w,x)$ and $g_0(w,x)$ for this case
following from our calculations are shown in Fig. \ref{fig1} for scaled rapidities $w=0,1,3,5,7$.
(with $\bar{\alpha}=0.2$ this corresponds to natural rapidities 0,5,15,25 and 35).

\begin{figure}
\begin{center}
\epsfig{file=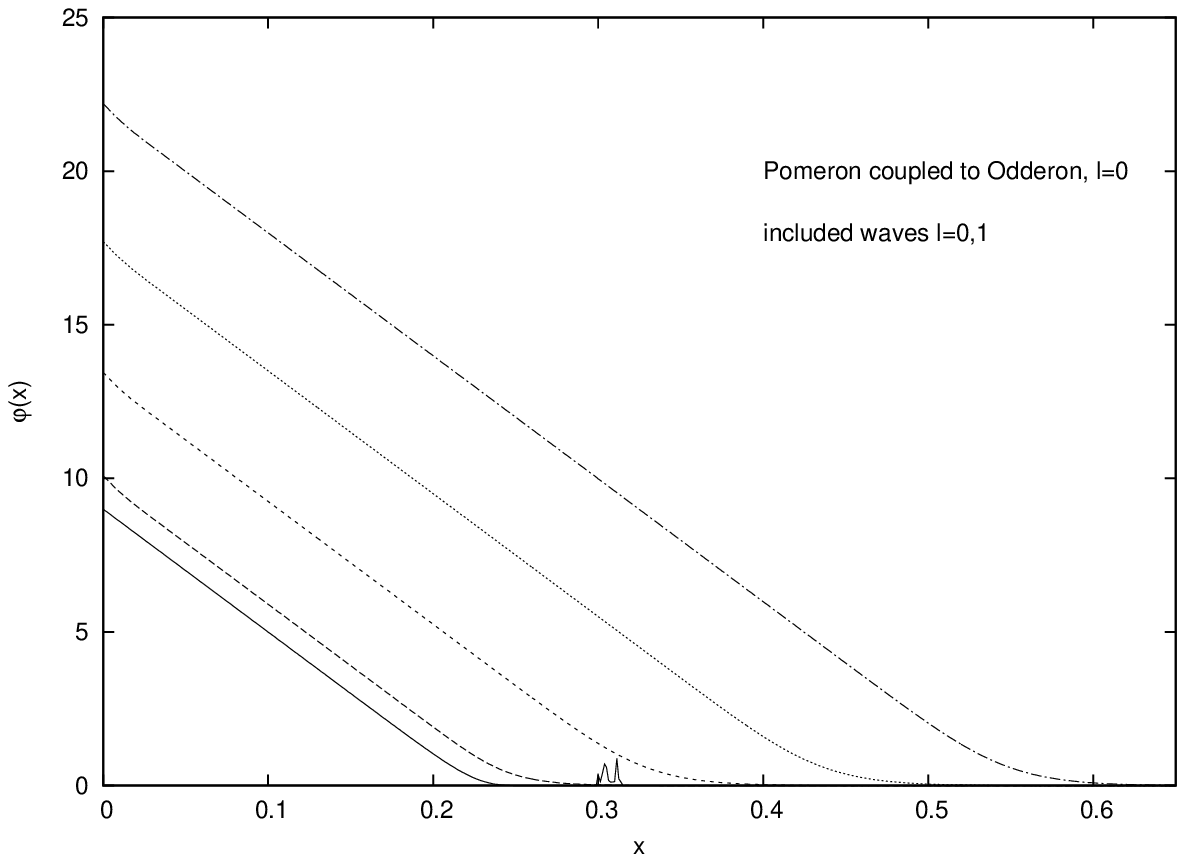, width=0.45\columnwidth }
\epsfig{file=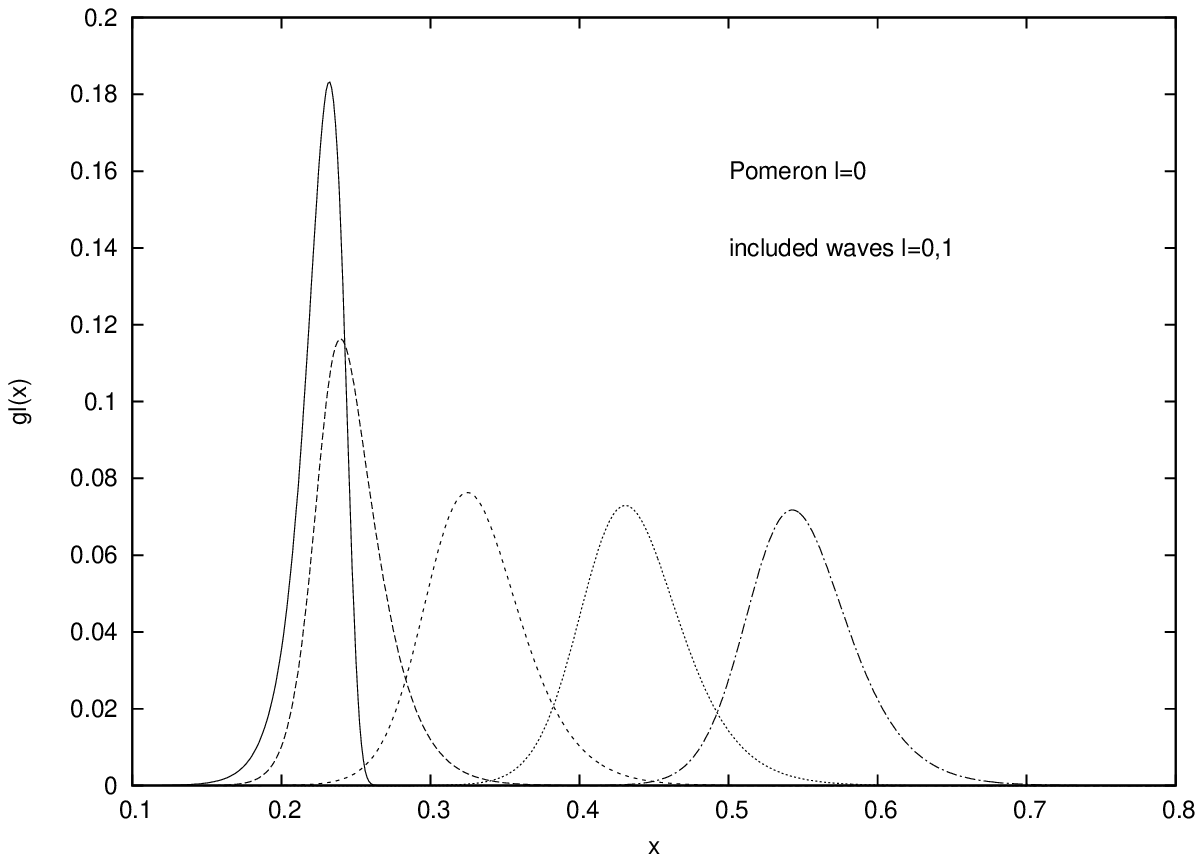, width=0.45\columnwidth }
\end{center}
\caption{The Pomeron $\phi_0$ coupled to  odderon $\psi_0$ (left panel) and the corresponding gluon density
(right panel). Curves from top to bottom in the left panel and from left to right in the right panel
correspond to $w=0,1,3,5$ and 7 }
\label{fig2}
\end{figure}

Now we take into account the  coupling of the pomeron and odderon on the minimal level, introducing all nonlinear terms
different from zero but restricting the partial waves for both $\phi$ and $\psi$ to $l=0$ (that is taking the pomeron at $L=0$ and odderon at $L=1$)
Our results are shown in Fig. \ref{fig2}
As one observes the change due to coupling with the odderon is barely visible.

\begin{figure}
\begin{center}
\epsfig{file=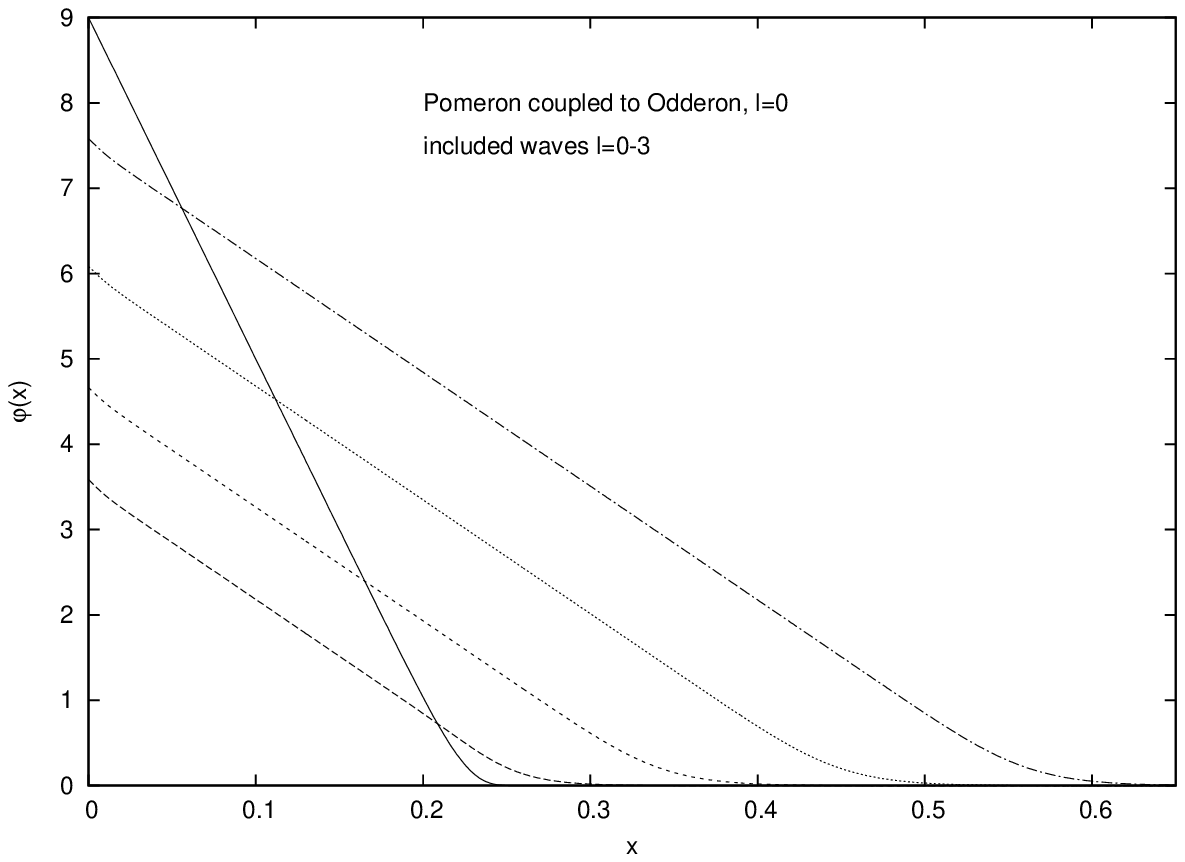, width=0.45\columnwidth }
\epsfig{file=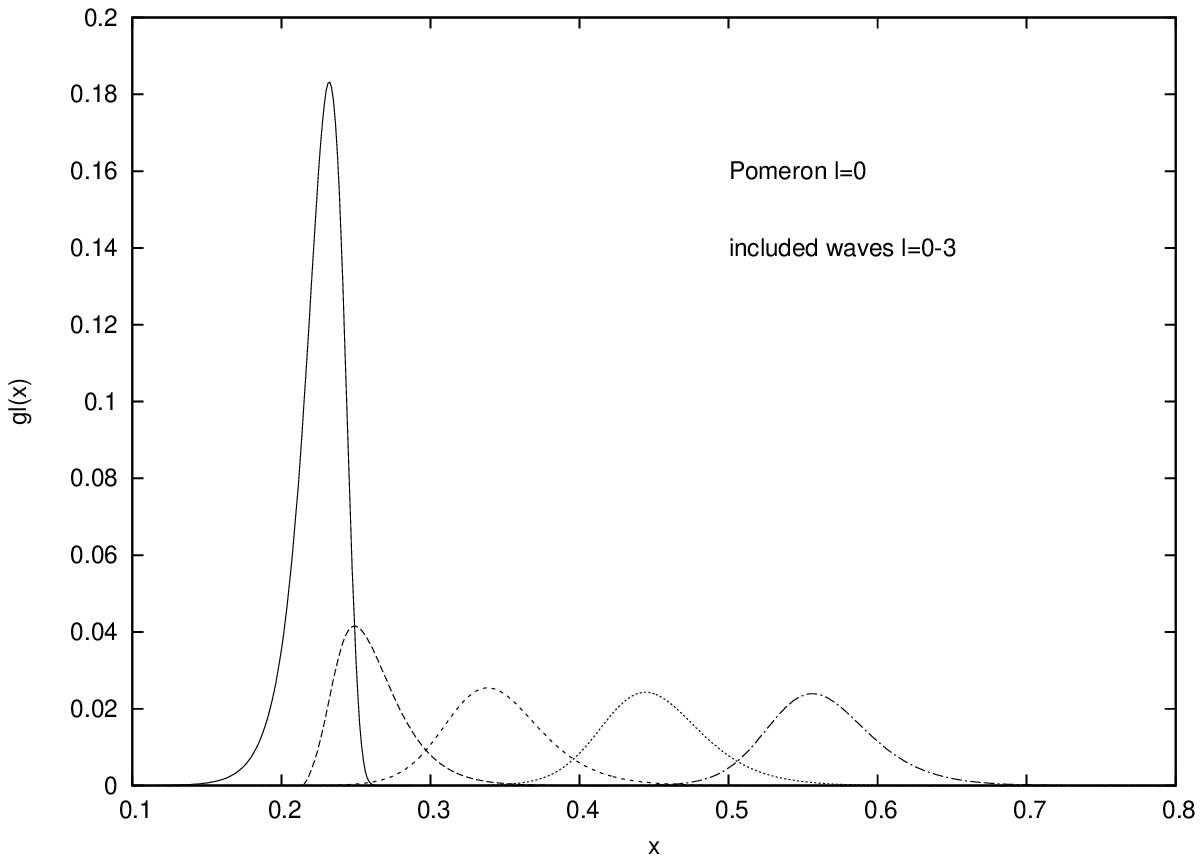, width=0.45\columnwidth }
\end{center}
\caption{The Pomeron $\phi_{0}$ coupled to the pomeron $\phi_1$ and  odderons $\psi_{0,1}$ (left panel) and the corresponding gluon density
(right panel). Curves from top to bottom in the left panel and from left to right in the right panel
correspond to $w=0,1,3,5$ and 7 }
\label{fig3}
\end{figure}

At the next step we widen the set of partial waves to include values $l=0$ and $l=1$, This implies taking into account also
the pomeron with $L=2$ and odderon with $L=3$. Calculation give the results shown in Fig. \ref{fig3}.
As we observe inclusion of  higher partial wave has a quite large influence on the evolution of the normal
pomeron with $L=0$. Already at $w=1$ both its amplitude and gluon density become more than twice reduced, although the general\
behavior with the growth of rapidity remains the same.

\begin{figure}
\begin{center}
\epsfig{file=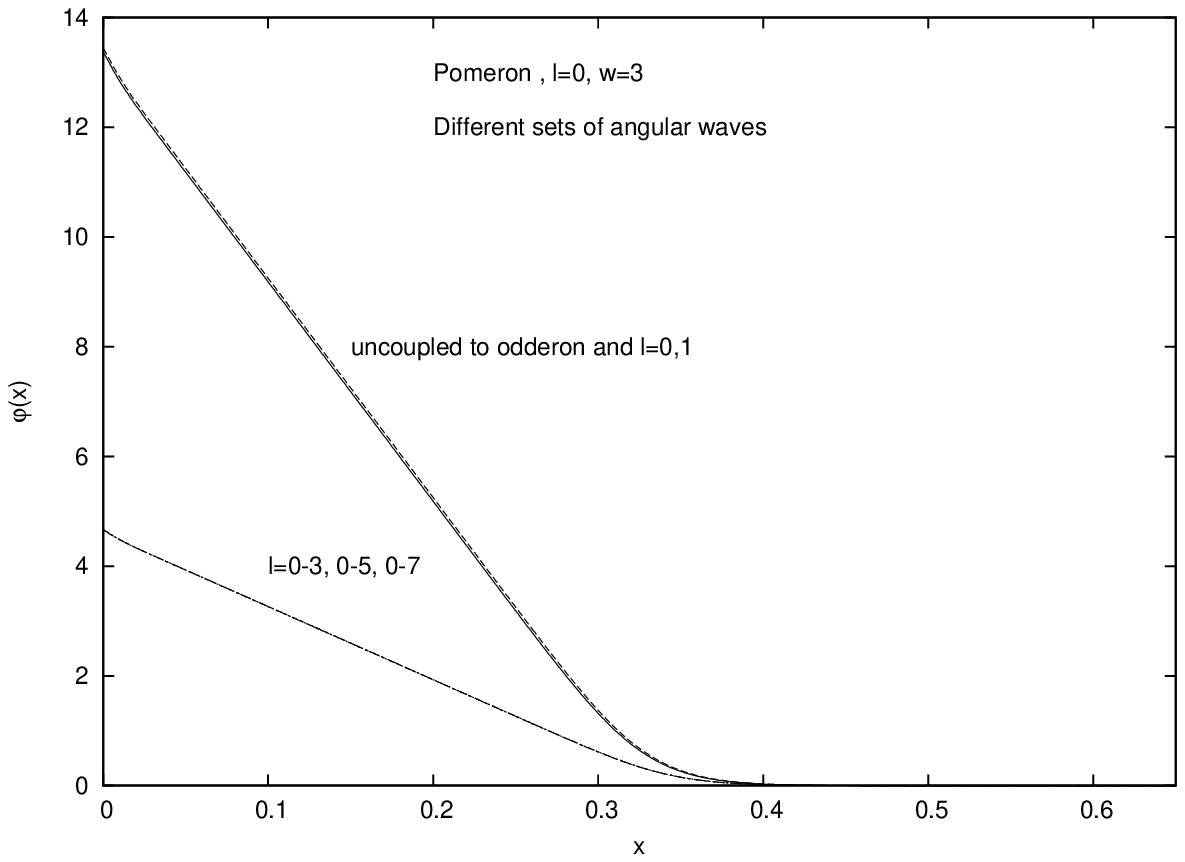, width=0.45\columnwidth }
\epsfig{file=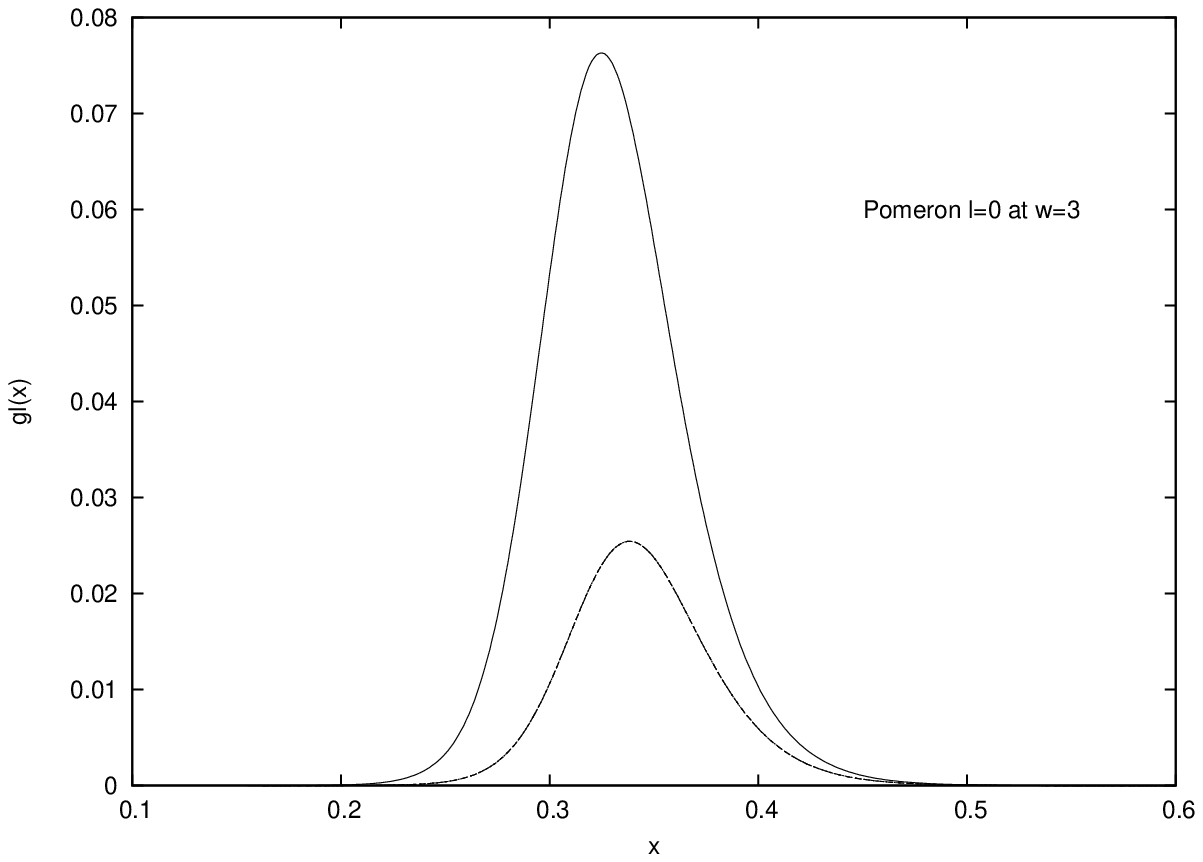, width=0.45\columnwidth }
\end{center}
\caption{The Pomeron $\phi_{0}$  (left panel) and the corresponding gluon density
(right panel) at $w=3$ and for different sets of included waves.
The upper curves  in both panel correspond to uncoupling and waves restricted to $l=0$
The lower curves corresponds to sets of waves (0,1),(0-2) and (0-3)}
\label{fig4}
\end{figure}

\begin{figure}
\begin{center}
\epsfig{file=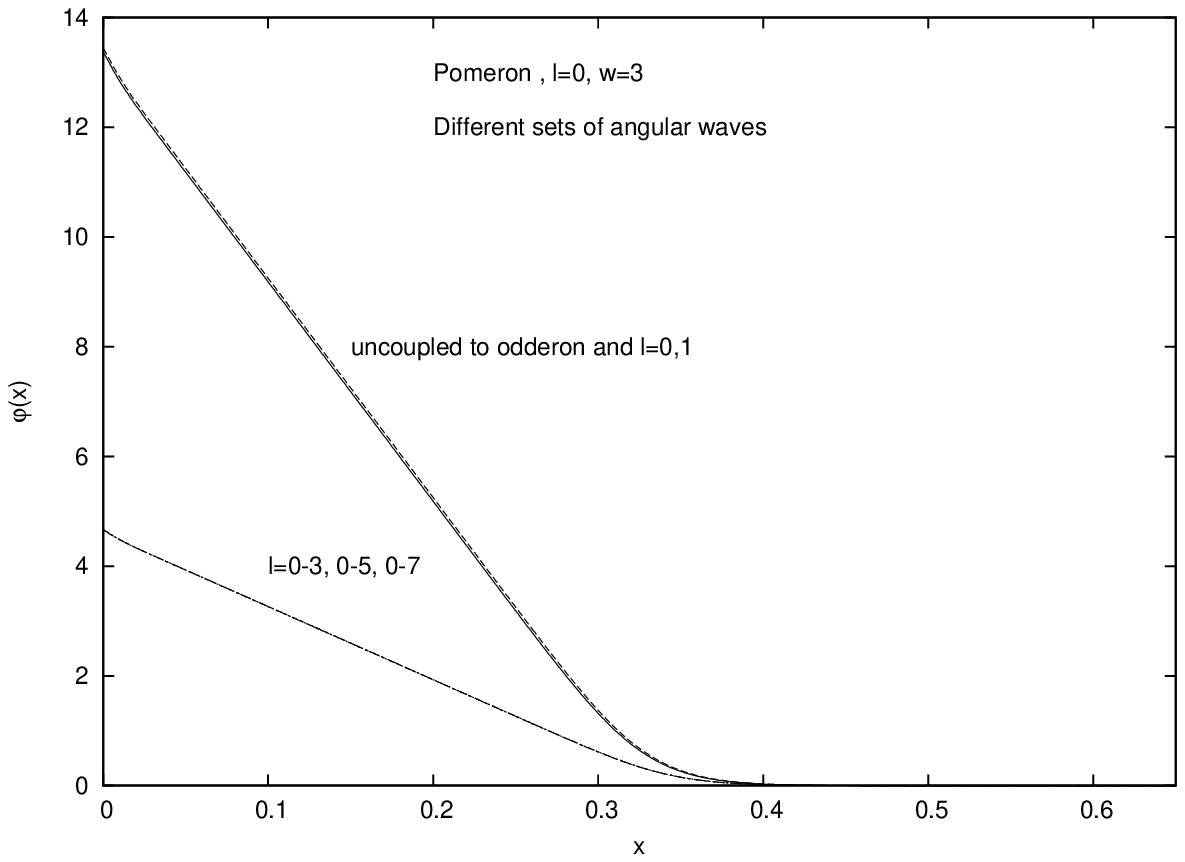, width=0.45\columnwidth }
\epsfig{file=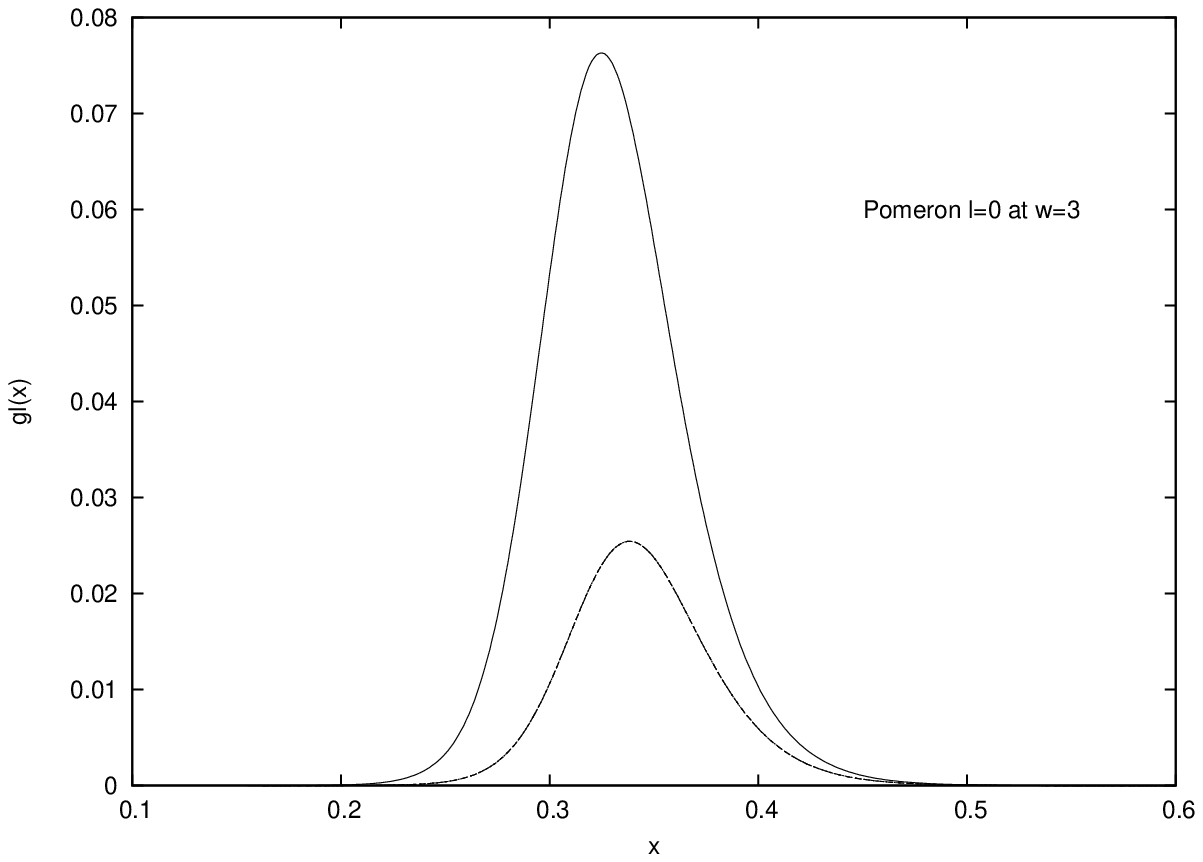, width=0.45\columnwidth }
\end{center}
\caption{The Pomeron $\phi_{0}$  (left panel) and the corresponding gluon density
(right panel) at $w=7$ and for different sets of included waves.
The upper curves  in both panel correspond to uncoupling and waves restricted to $l=0$
The lower curves corresponds to sets of waves (0,1),(0-2) and (0-3)}
\label{fig5}
\end{figure}

Remarkably inclusion of more partial waves does not change the pomeron $\phi_0$. This is illustrated in
Figs. \ref{fig4} and \ref{fig5} where we show $\phi_0$ and $g_0$ at $w=3$ and $w=7$ respectively for different
sets of included waves:(0),(0,1),(0-2) and (0-3) together with the uncoupled case.

\begin{figure}
\begin{center}
\epsfig{file=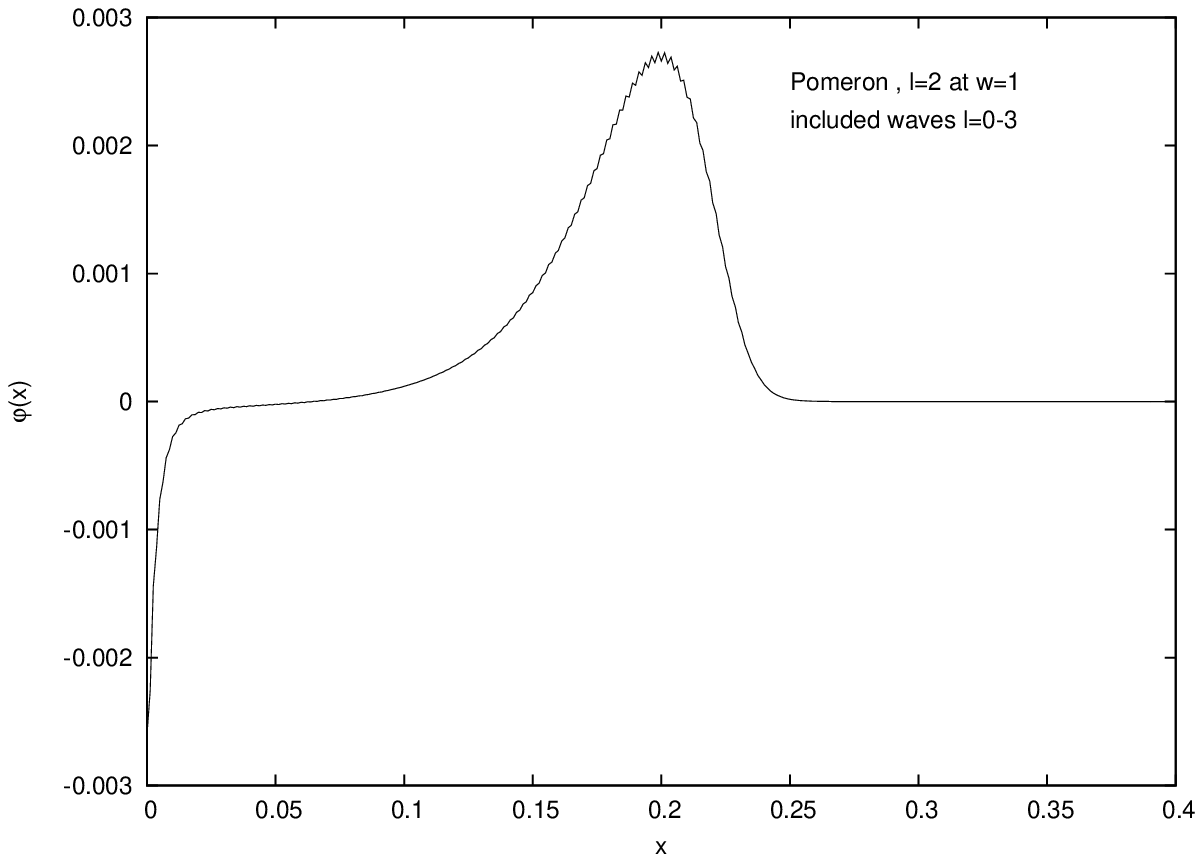, width=0.45\columnwidth }
\epsfig{file=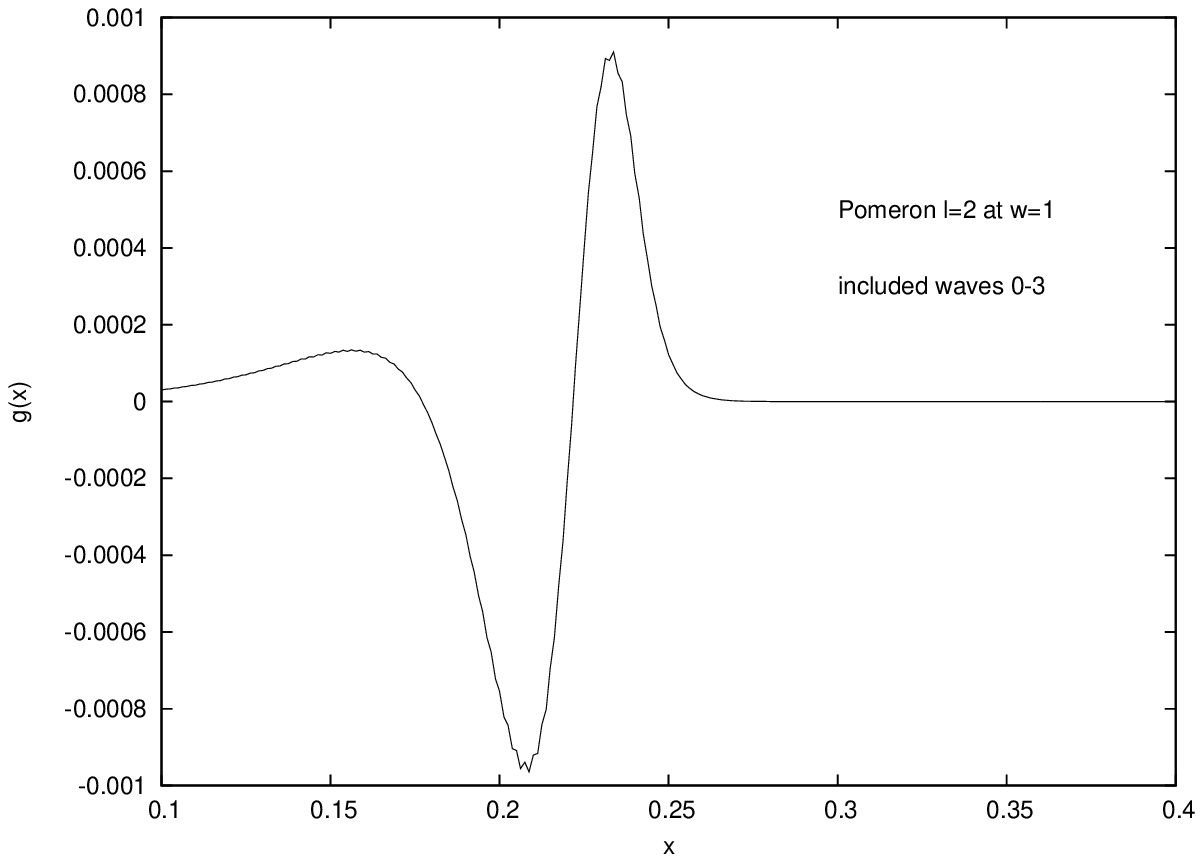, width=0.45\columnwidth }
\end{center}
\caption{The Pomeron $\phi_{1}$  (left panel) and the corresponding "gluon density"
(right panel) at $w=1$ with included waves 0,1.}
\label{fig6}
\end{figure}

Inclusion of higher partial waves leads to appearance of amplitudes $\phi_{1,2,3}$ which
correspond to pomerons with $L=2,4,6$, which result from the evolution in the presence of the
external direction, although they are zero initially. These amplitudes are small and rapidly
diminish with the rapidity in accordance with their behavior under the BFKL evolution.
Starting from $w=3$ they are all practically equal to zero. They are also practically independent
from the inclusion of higher partial waves. So we illustrate them only at the
earlier part of the evolution at $w=1$ and $w=2$ and for the minimal sets of partial waves.
In Fig. \ref{fig6} we show $\phi_1$ at $w=1$ with included waves 0,1 and the corresponding
"gluon density", that is its double derivative in $t$
As one observes $\partial_t^2\phi(t)$ is not positive and can hardly be interpreted as
"density".

In the next figure we illustrate $\phi_1$ at $w=2$ and $\phi_2$ at $w=1$
For $\phi_1$ waves with $l=0,1$ are included, for $\phi_2$ waves from $l=0$ to $l=2$ are included.
\begin{figure}
\begin{center}
\epsfig{file=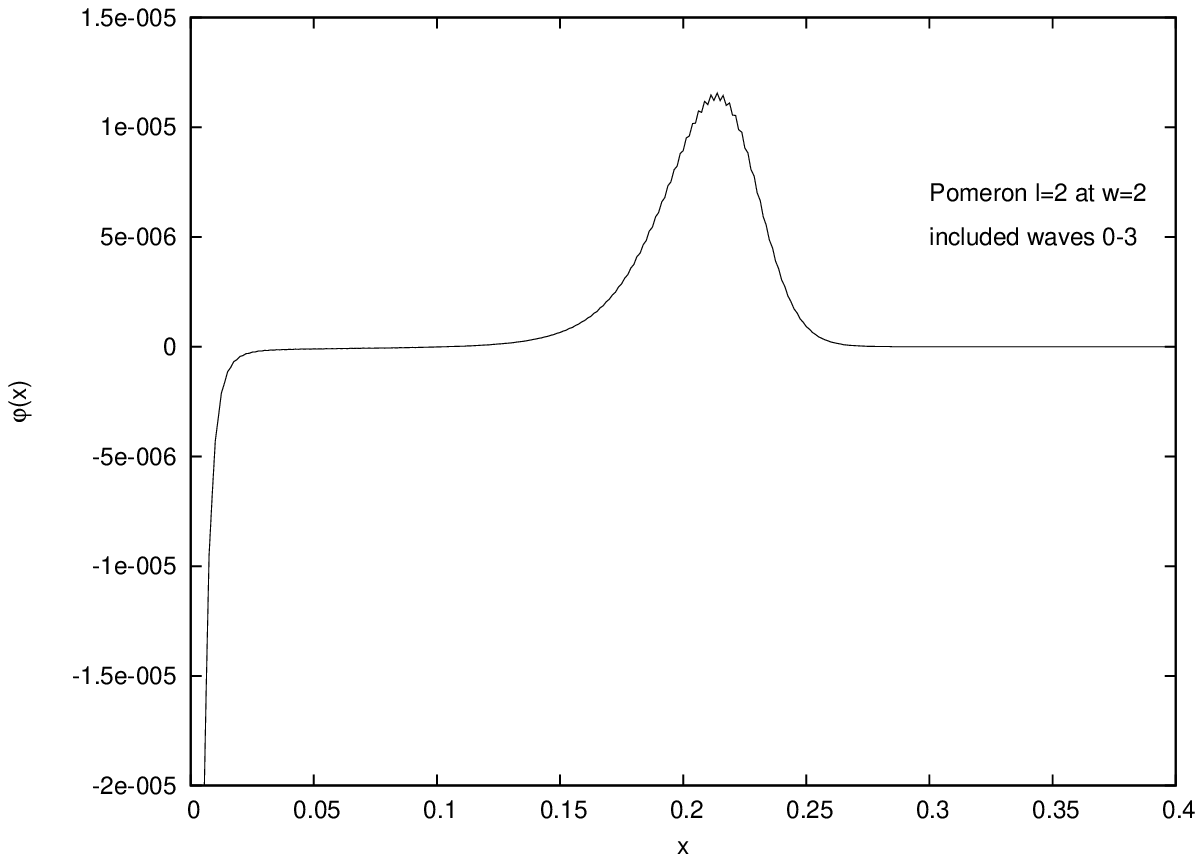, width=0.45\columnwidth }
\epsfig{file=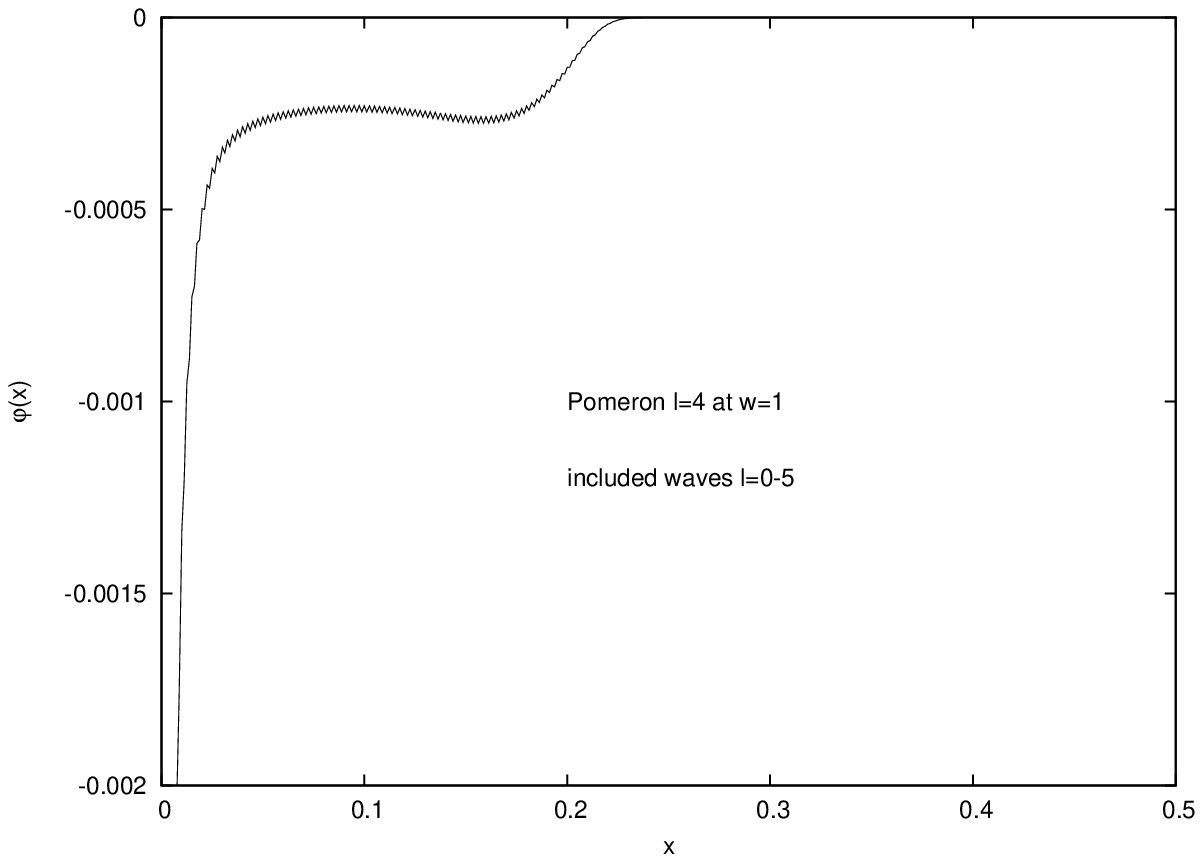, width=0.45\columnwidth }
\end{center}
\caption{The Pomerons $\phi_{1}$  at $w=2$(left panel) and $\phi_2$ at $w=1$
(right panel) at  with included waves l=0,1 for $\phi_1$ and $l=0,1,2$ for $\phi-2$}
\label{fig7}
\end{figure}

\subsection{The odderon}
\begin{figure}
\begin{center}
\epsfig{file=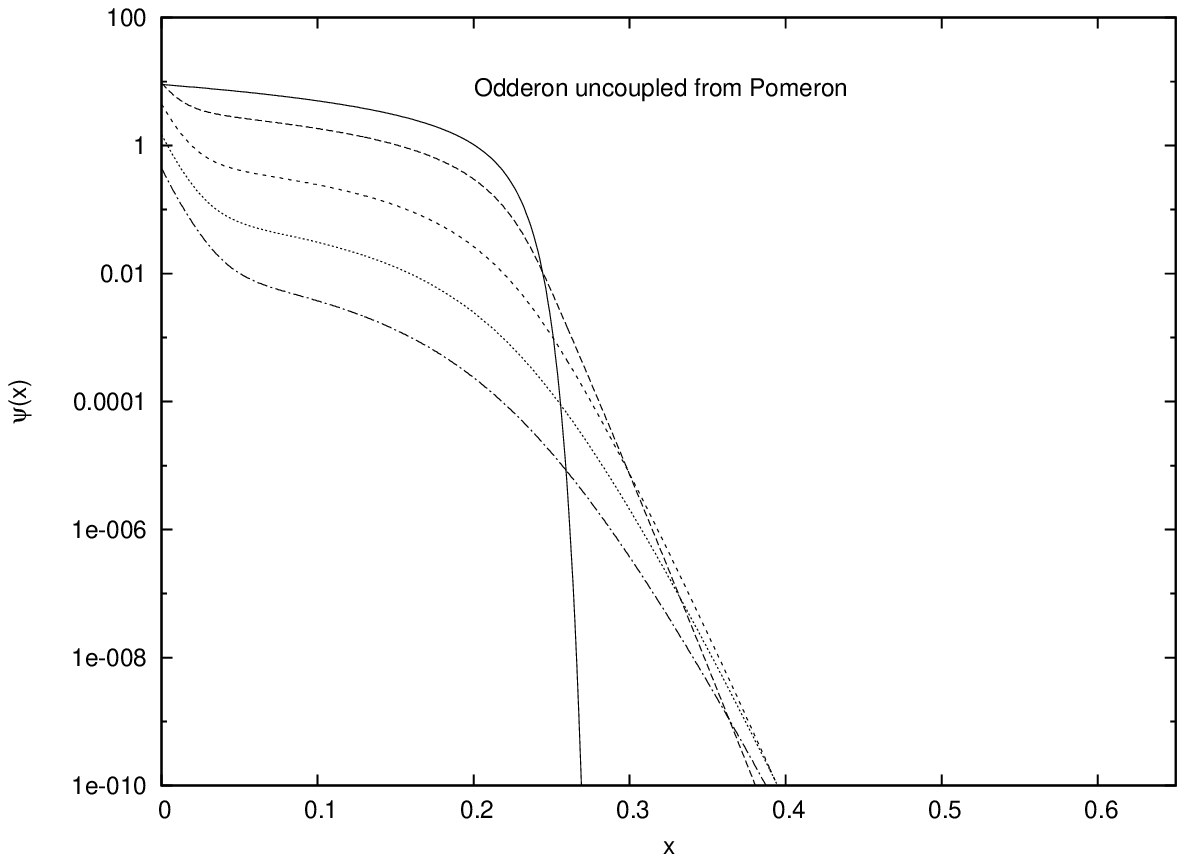, width=0.45\columnwidth }
\epsfig{file=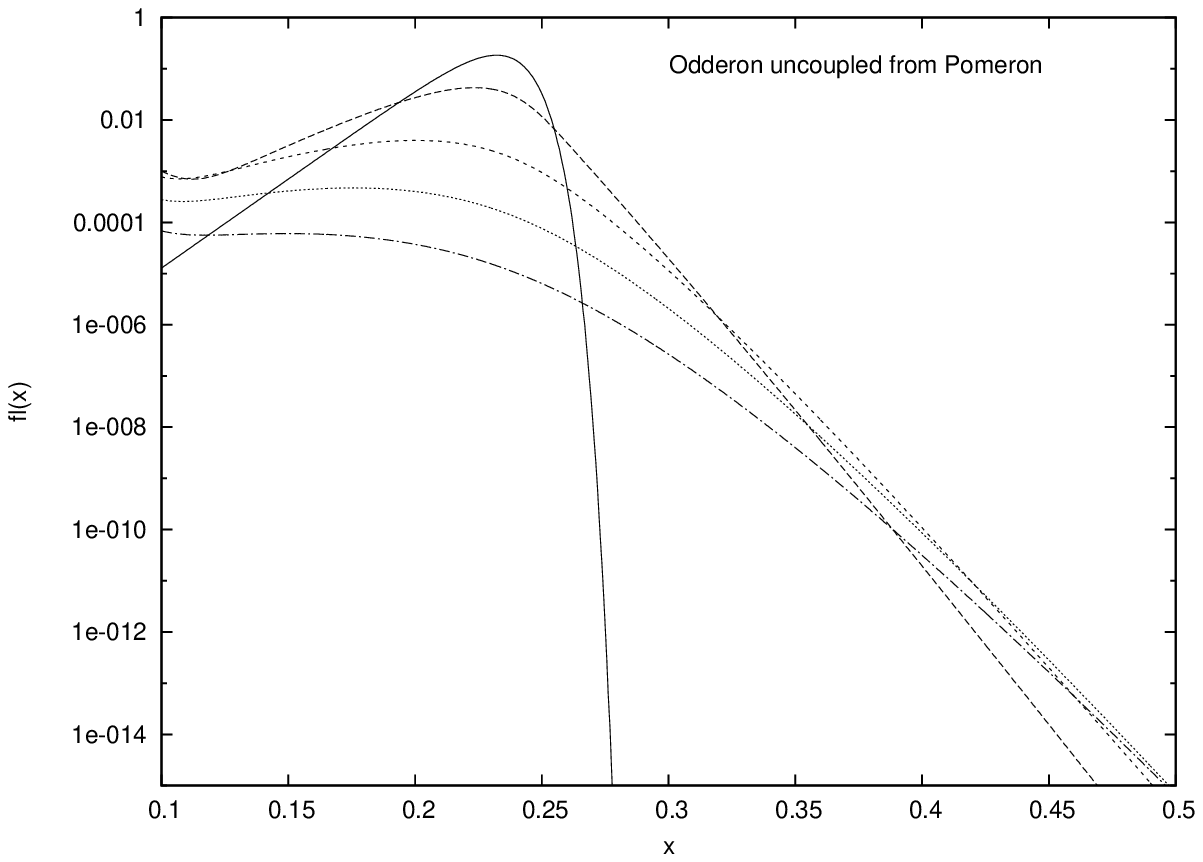, width=0.45\columnwidth }
\end{center}
\caption{The Odderon $\psi_0$ decoupled from the pomeron (left panel) and the corresponding "gluon density"
(right panel). Curves from top to bottom in both panels
correspond to $w=0,1,3,5$ and 7 }
\label{fig8}
\end{figure}

Again we start with the situation when the odderon and pomeron are decoupled.
We assume that it is the "normal" odderon with $L=1$ , corresponding to $\psi_0$, which
interacts with the target. Then all odderons with higher angular momenta will be zero and
$\psi_0$ will evolve according to the BFKL Hamiltonian. In this case our calculations give the results shown in
Fig. \ref{fig8}, where apart from $\psi_0$ we illustrate also the corresponding "gluon density"
$f_0=\partial_t^2\psi_0$ (as we mentioned its physical interpretation is rather obscure). As for the pomeron we show
both $\psi_0$ and $f_0$ as functions of $x$ for rising rapidities $w=0,1,3,5$ and 7. Since the odderon functions
go to zero with the growth of momentum very fast we present these and the following results in the logarithmic scale.

\begin{figure}
\begin{center}
\epsfig{file=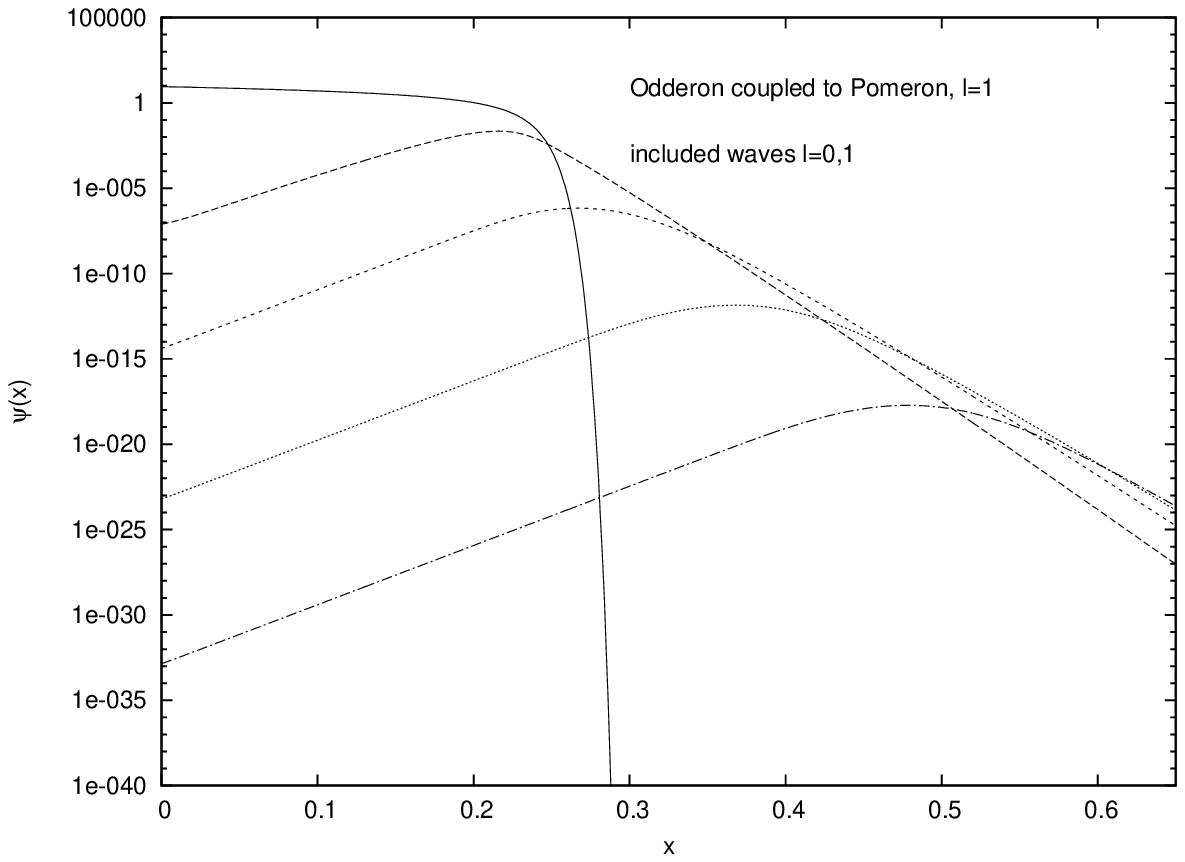, width =0.45\columnwidth }
\epsfig{file=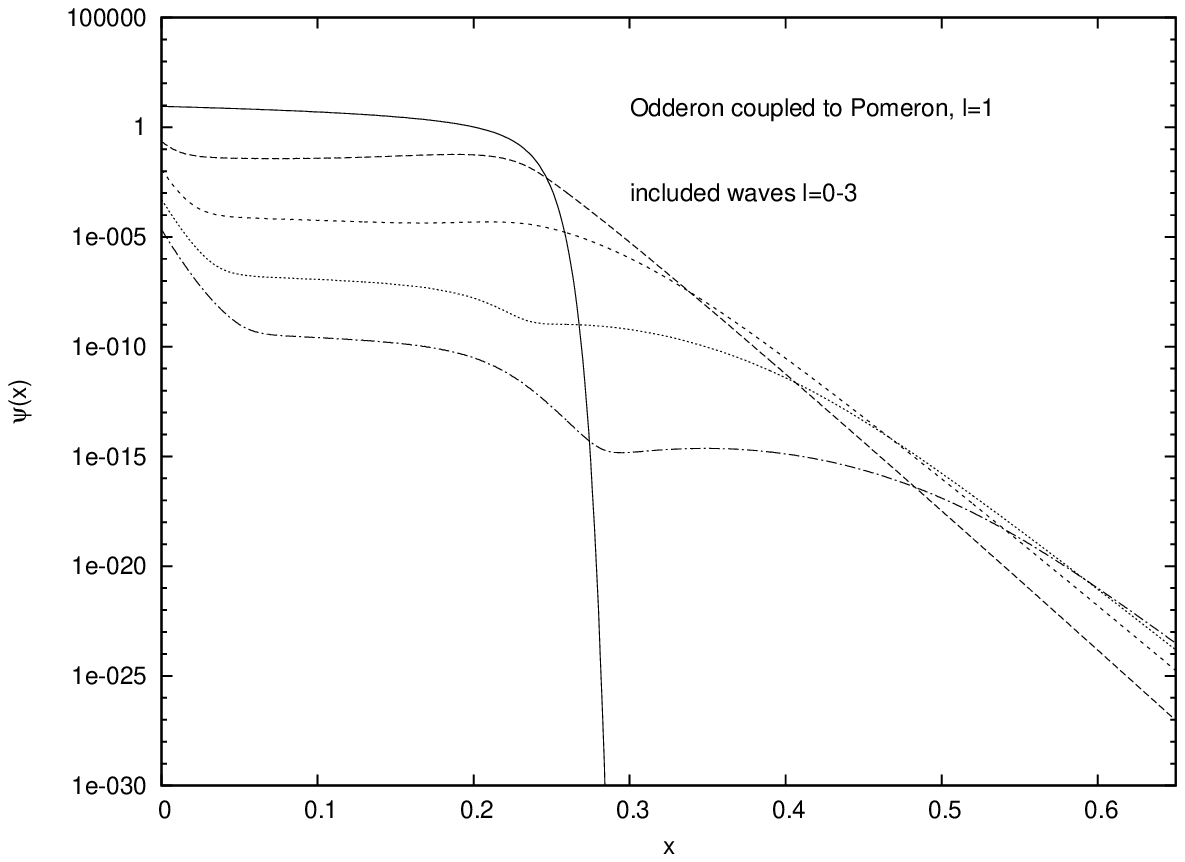, width=0.45\columnwidth }
\end{center}
\caption{The Odderon $\psi_0$ with waves $l=0,1$ included (left panel) waves $l=0-3$ included
(right panel). Curves from top to bottom in both panels
correspond to $w=0,1,3,5$ and 7 }
\label{fig9}
\end{figure}

Coupling to the pomeron leads to strong reduction of odderon amplitudes. It is particulary strong if only
waves 0 and 1 are taken into account. If the waves include $l=0-3$ the reduction is weaker but still persists. This
is illustrated in Fig. \ref{fig9} in which $\psi_1$ is shown for both sets of waves.

\begin{figure}
\begin{center}
\epsfig{file=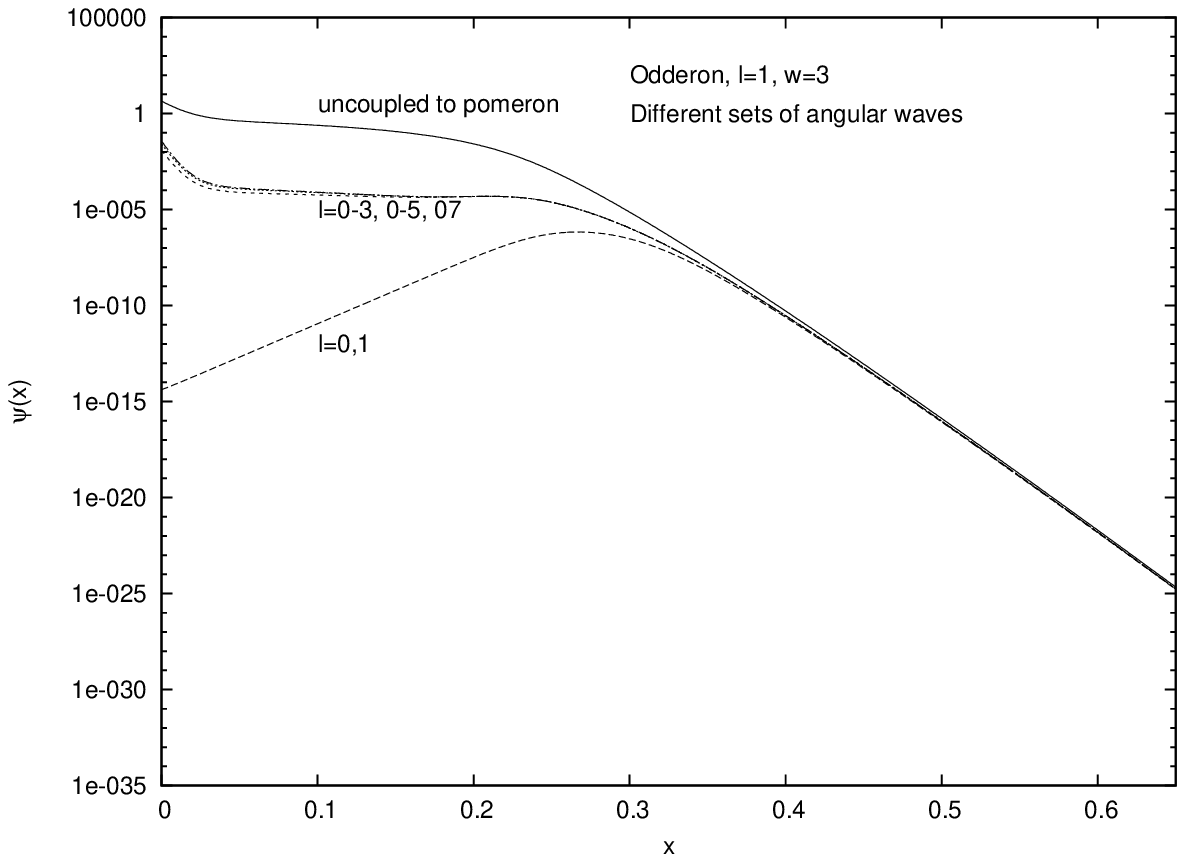, width=0.45\columnwidth }
\epsfig{file=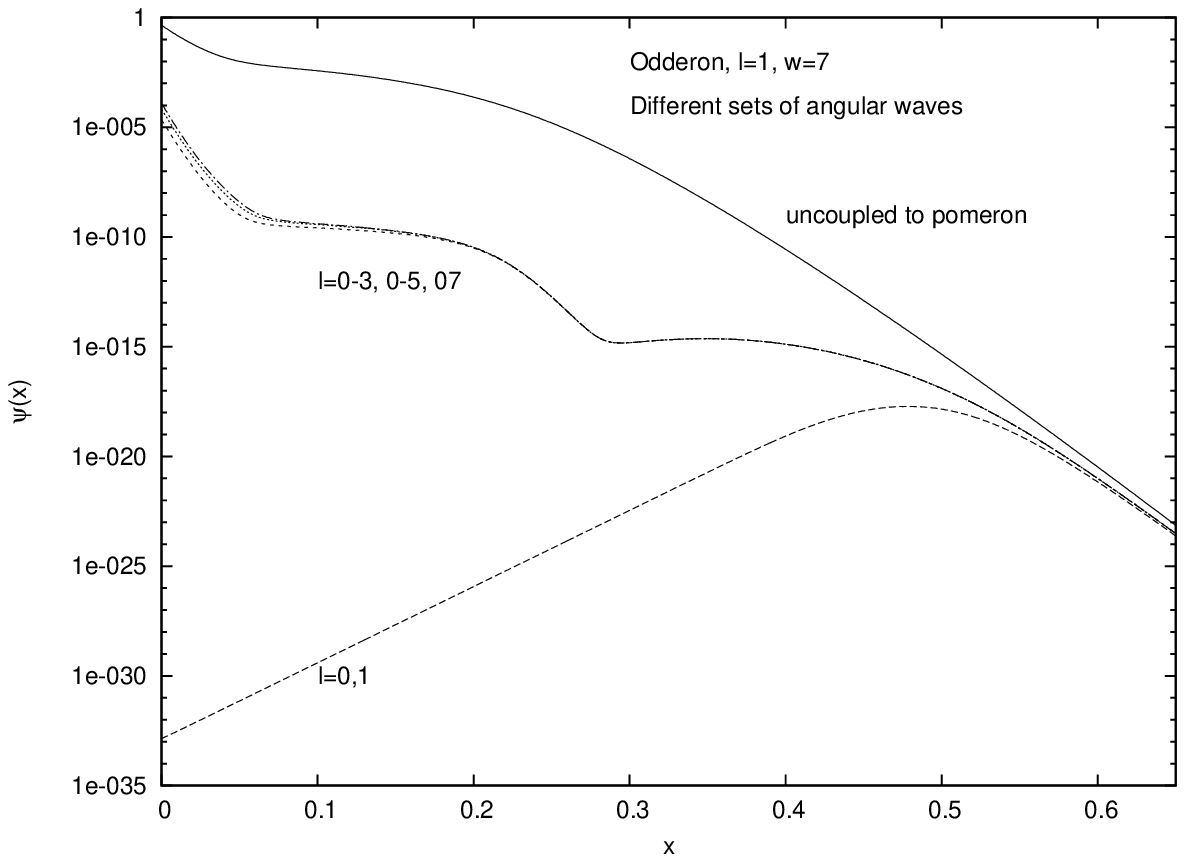, width=0.45\columnwidth }
\end{center}
\caption{The Odderon $\psi_0$ at $w=3$(left panel) and $w=7$  (right panel)
with different sets of partial waves included.}
\label{fig10}
\end{figure}

As with the pomeron, further widening of the set of partial waves does not change the odderon function
$\psi_1$. This is illustrated in Fig. \ref{fig10} where we plot $\Psi_1$ at $w=3$ and $w=7$ for different sets
of partial waves.

As to "gluon density" $f=\partial_t^2\psi$ it turns out to be very small, goes to zero very fast with rapidity
and changes sign. In Fig. 11 we show it in the logarithmic scale at all rapidities and also in the normal scale
at $w=1$. Use of the logarithmic scale leads to breaks in the curves at intervals where $f<0$
\begin{figure}
\begin{center}
\epsfig{file=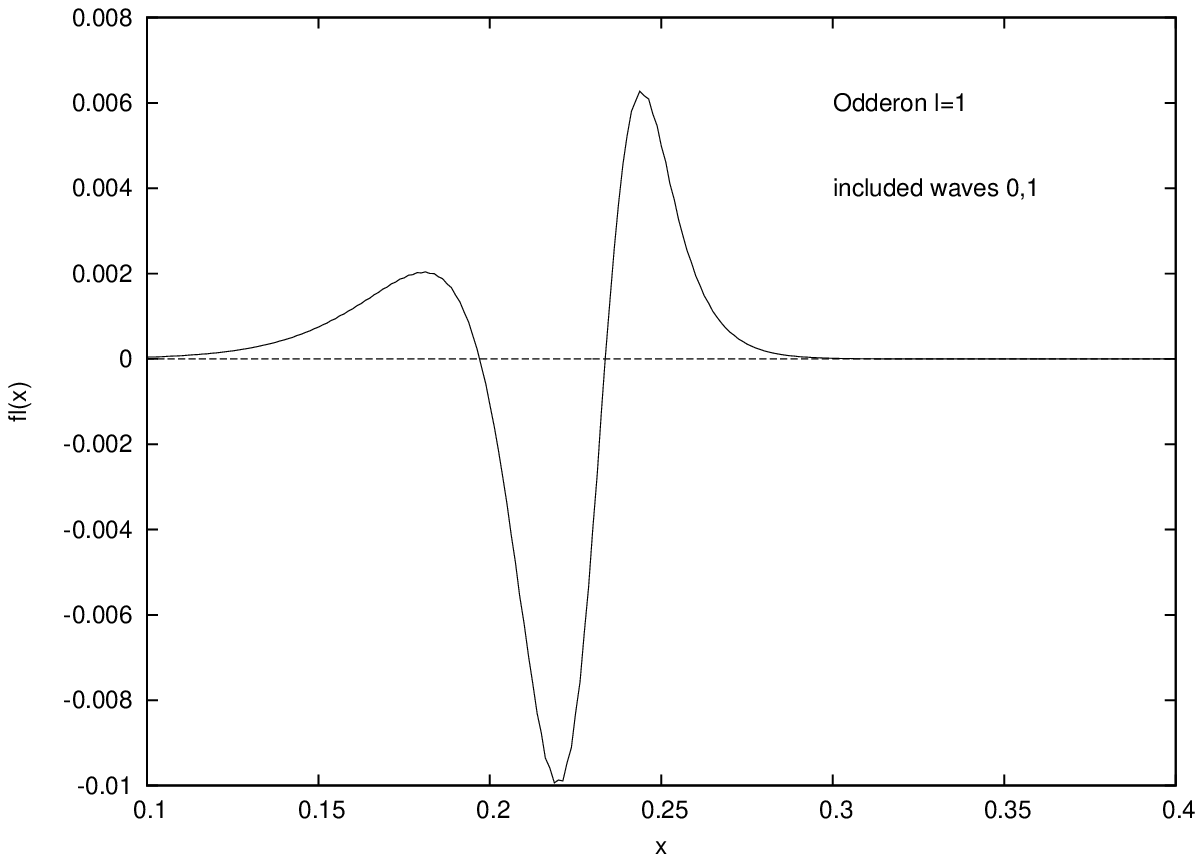, width=0.45\columnwidth }
\epsfig{file=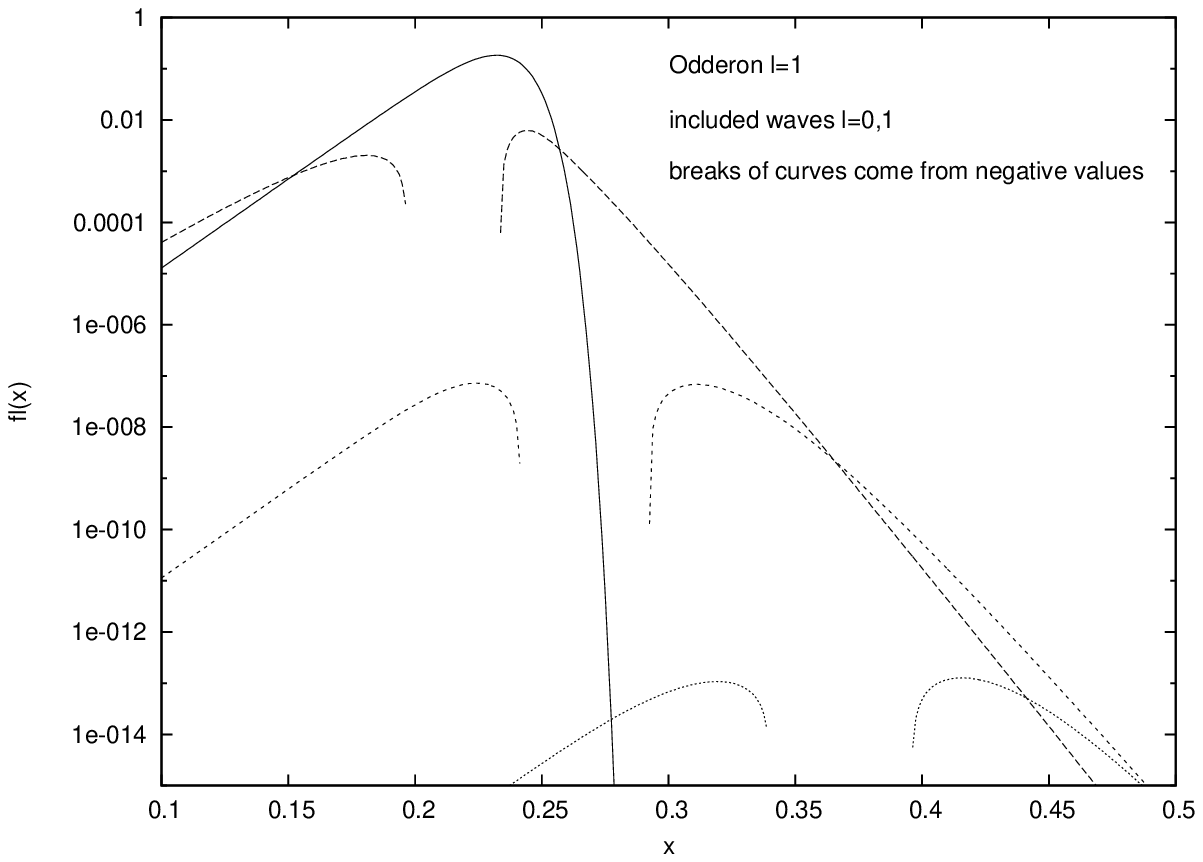, width=0.45\columnwidth }
\end{center}
\caption{The "gluon density" $f_0= \partial_t^2\psi$ at $w=1$ (left panel) and $w=0,1,3,5$  (right panel, from top down)
with partial waves included $l=0,1$.}
\label{fig11}
\end{figure}

\begin{figure}
\begin{center}
\epsfig{file=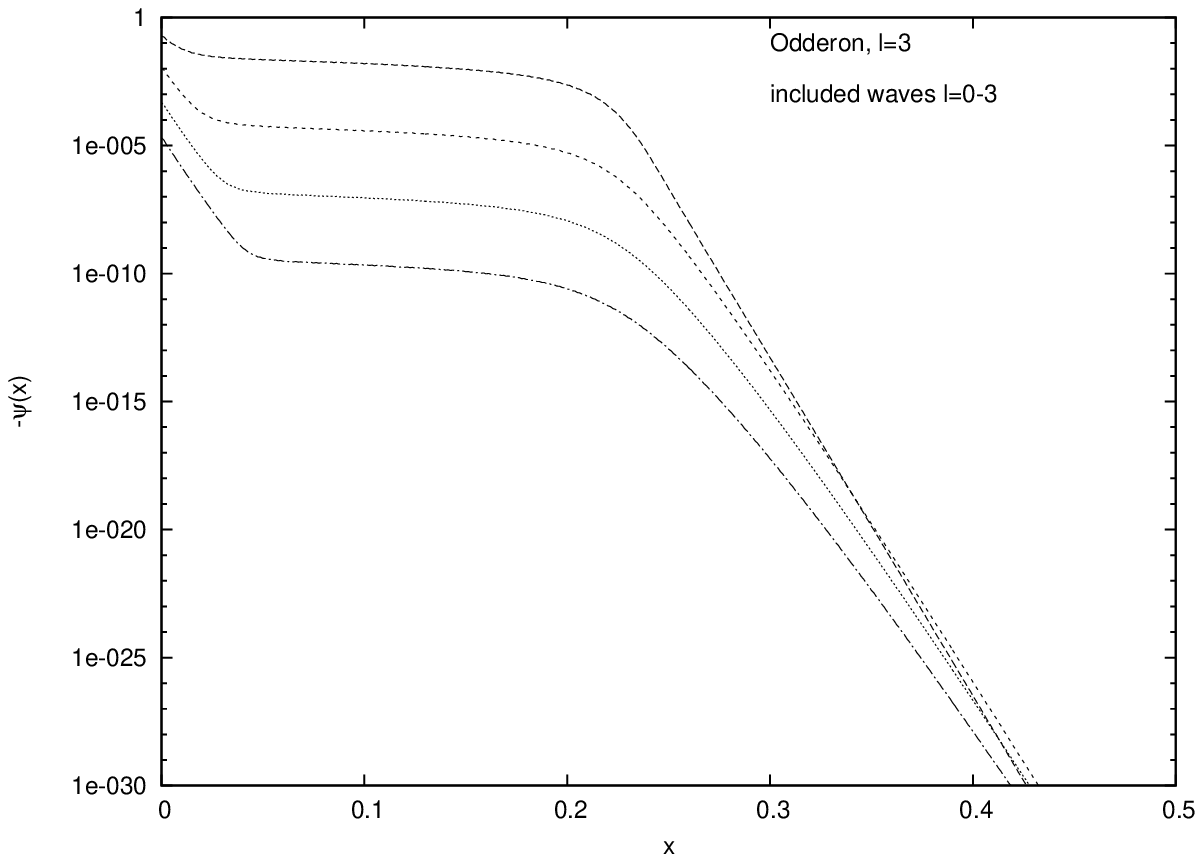, width=0.45\columnwidth }
\epsfig{file=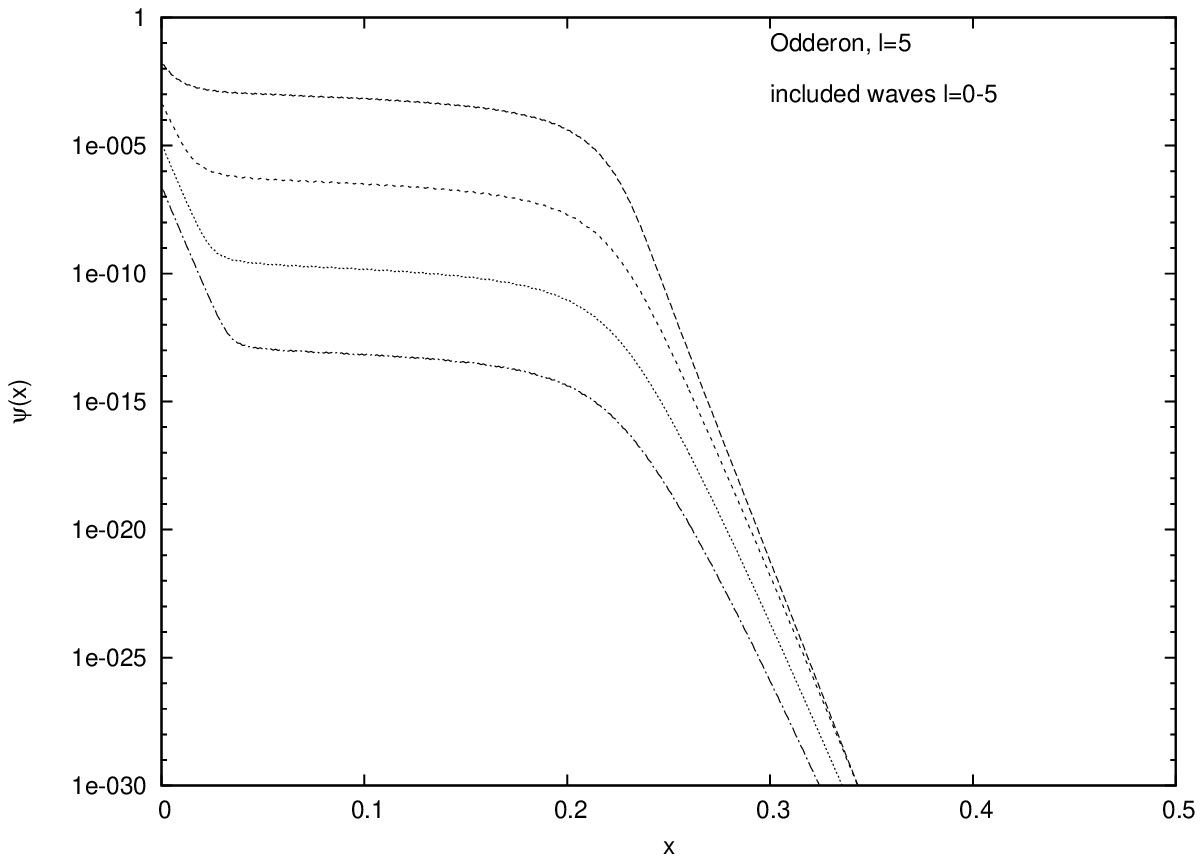, width=0.45\columnwidth }
\end{center}
\caption{The odderons with $L=3$ (left panel) and $L=5$ (right panel) at $w=1,3,5,7$ from top down}
\label{fig12}
\end{figure}

We finally come to the odderons with higher $l$. They go down quickly with the growth of $l$ and their behavior with $x$ and rapidity is similar
to the odderon with $L=1$. In Fig. \ref{fig12} we show the odderons with $L=3$ and $L=5$.

\begin{figure}
\begin{center}
\epsfig{file=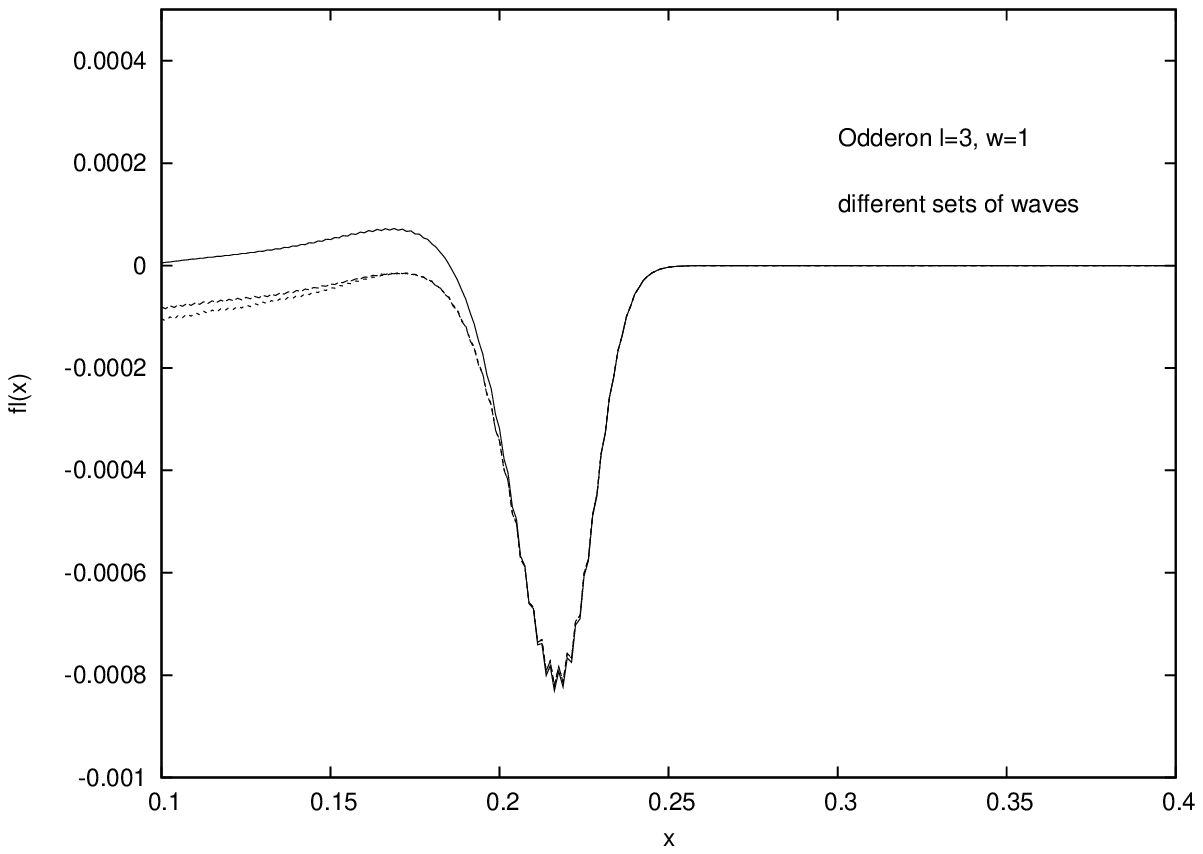, width=0.45\columnwidth }
\epsfig{file=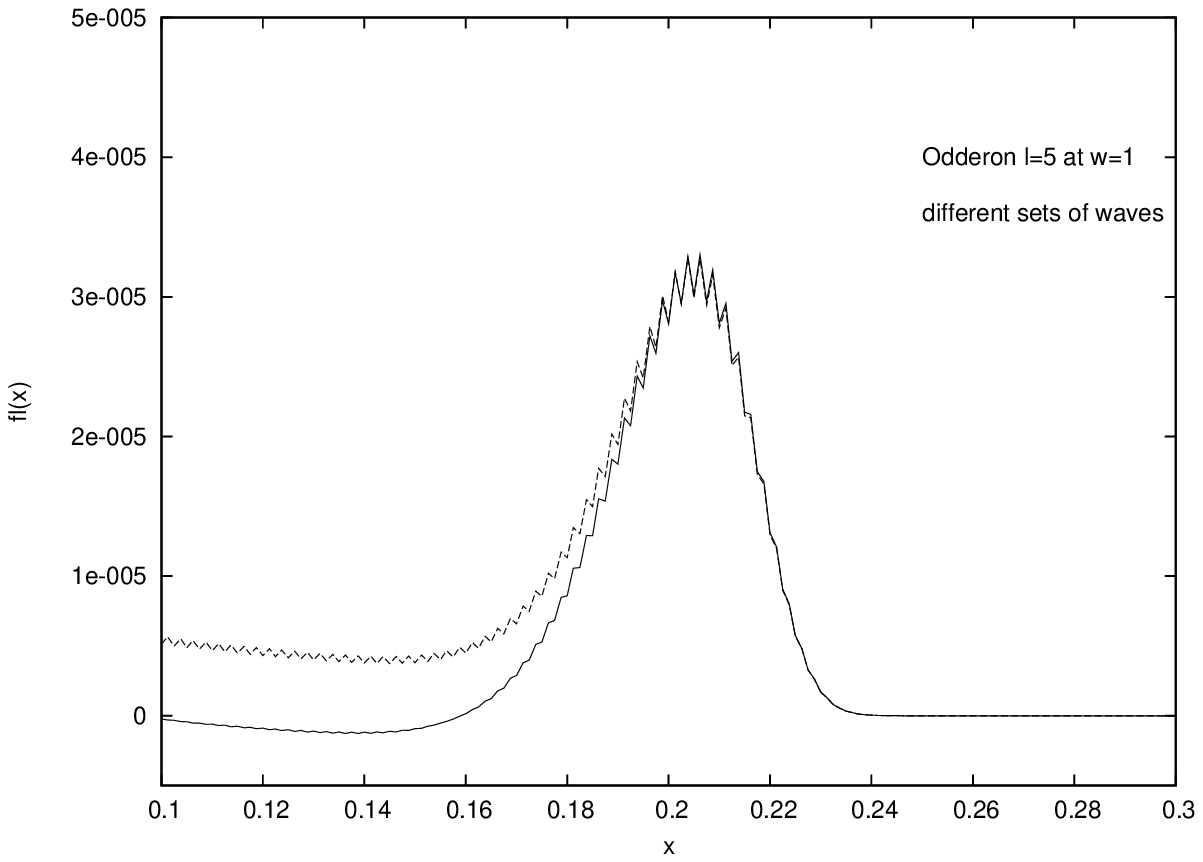, width=0.45\columnwidth }
\end{center}
\caption{The "gluon densities" $f_1= \partial_t^2\psi_1$ (left panel) and
 $f_2= \partial_t^2\psi_2$ (right panel) at $w=1$ }
\label{fig13}
\end{figure}

The corresponding "gluon densities" $f_1$ and $f_2$ at $w=1$ are presented in Fig. \ref{fig13}
At larger $w$ all $f_l$ with $l\geq 1$ are extremely small.

\section{Discussion}
We have studied the system of coupled evolution equations for the pomeron and odderon, derived in
~\cite{ksw, hiim} in the tranlationally invariant approximation proposed in ~\cite{motyka}
taking in account the full angular dependence. Our numerical calculations on the whole confirm qualitative predictions
made in these references about the strong damping of the odderon and its fast diminishing with the growing
rapidity as a result of its interaction with the pomeron.
In our calculations  we discovered that the inverse influence on the pomeron of the interaction with the odderon also damps both the amplitude and the gluon density.
as soon as one goes beyond the
basic pomeron and odderon states with $L=0$ and $l=1$.  With only the basic odderon state $L=1$ the pomeron practically does not change. But inclusion of higher states,
starting from the pomeron at $L=2$ and odderon at $l=3$
substantially reduces the basic pomeron state at $l=0$ while preserving its qualitative dependence on rapidity and momentum. This reduction does not practically change with the
the number of states with $L>1$ included.
This may be considered as the main and somewhat unexpected result of our calculations with possible physical consequences.

In physical applications for the collision of azimuthal symmetric projectile with the target nucleus the angular dependence is obviously average out,
so that all partial waves go to zero expect at $L=0$. So the only surviving state is precisely the basic pomeron with $l=0$. Without the odderon
it evolves according to the BK equation. Our calculations show that as soon as one takes into account states with $L>1$, which appear in the course of the
evolution, both  pomeron amplitude and the corresponding gluon density turn out to be more than twice reduced. Of course this prediction has been made under the assumption
that the interaction of the odderon with the nucleon is of the  same magnitude as that of the pomeron. In perturbation theory the odderon interaction carries one
extra $\alpha_s$ and so is significantly smaller. However this interaction is in fact non-perturbative and its magnitude is unknown {\it apriori}.
One may hope that experimental observations may shed light on this question.

\end{document}